\documentclass[aps,prb,floatfix,amsmath,amssymb,eqsecnum,nofootinbib,twocolumn,superscriptaddress]{revtex4-1}
\usepackage[usenames,dvipsnames]{color}

\usepackage{graphicx,xcolor}
\usepackage{dcolumn}
\usepackage{bm}
\usepackage{amsmath}    
\usepackage{relsize}  

\begin{document}

\title{Buckling transitions and clock order of two-dimensional Coulomb crystals}

\author{Daniel Podolsky}
\affiliation{Department of Physics, Technion, Haifa 32000, Israel}

\author{Efrat Shimshoni}
\affiliation{Department of Physics, Bar-Ilan University, Ramat-Gan
52900, Israel}

\author{Giovanna Morigi}
\affiliation{Theoretische Physik, Universit\"{a}t des Saarlandes,
D-66123 Saarbr\"{u}cken, Germany}

\author{Shmuel Fishman}
\affiliation{Department of Physics, Technion, Haifa 32000, Israel}

\date{\today}

\begin{abstract}
Crystals of repulsively interacting ions in planar traps form
hexagonal lattices, which undergo a buckling instability towards a
multi-layer structure as the transverse trap frequency is reduced.
Numerical and experimental results indicate that the new structure
is composed of three planes, whose separation increases
continuously from zero. We study the effects of thermal and
quantum fluctuations by mapping this structural instability to the
six-state clock model.  A prominent implication of this
mapping is that at finite temperature, fluctuations split the
buckling instability into two thermal transitions,
accompanied by the appearance of an intermediate critical phase.
This phase is characterized by quasi-long-range order in the
spatial tripartite pattern. It is manifested by broadened Bragg
peaks at new wave vectors, whose line-shape provides a direct
measurement of the temperature dependent exponent $\eta(T)$
characteristic of the power-law correlations in the
critical phase.  A quantum phase transition is found at the
largest value of the critical transverse frequency: here the
critical intermediate phase shrinks to zero.  Moreover, within the
ordered phase, we predict a crossover from classical to quantum
behavior, signifying the emergence of an additional characteristic
scale for clock order.  We discuss experimental realizations
with trapped ions and polarized dipolar gases, and propose that
within accessible technology, such experiments can provide a
direct probe of the rich phase diagram of the quantum clock model,
not easily observable in condensed matter analogues. Therefore, this works
highlights the potential for ionic and dipolar systems to serve as simulators for complex
models in statistical mechanics and condensed matter physics.
\end{abstract}

\maketitle


\section{Introduction}

The statistical physics of two dimensional systems is fascinating,
as it exhibits phases that are critical for a non-vanishing range
of parameters. In these phases there is no true long range order
and correlations decay as a power law. The pioneering work of
Kosterlitz and Thouless \cite{KosterlitzThouless} and of
Berezinskii \cite{Berezinskii} laid the foundation for the
theoretical analysis of such systems. This critical behavior
requires the symmetry of the Hamiltonian with respect to rotation
of the order parameter by some angle variable $\theta$. Its
physics is therefore captured by an easy-plane spin Hamiltonian
known as the $XY$-model.

An interesting situation may be encountered if the planar rotation
symmetry is broken by an additional term in the Hamiltonian, which
favors specific evenly-spaced values of the angle $\theta_i=2\pi i/q$, for $i=1,2...q$. The resulting
model is dubbed the $q$-state ($Z_q$)
clock model.  At low enough temperatures, the phases get pinned at one of the favored discrete values $\theta_i$, signaling the onset of true long range order.  Interestingly, for $q>4$ the thermal transition between ordered and disordered phases is separated by an intermediate phase with critical correlations\cite{Jose}.  At zero temperature, quantum fluctuations take over.  Then, the ordered and disordered phases are directly connected by a quantum critical point.\cite{Hove,Blankschtein,Oshikawa,Lou}

The unique features of the XY and clock models in 2D are manifested in a broad and expanding set of distinct physical systems. The most familiar realizations of the 2D XY-model are associated
with superfluidity and superconductivity in thin films, where several predictions
were verified experimentally
\cite{ChaikinLubensky,bishop,HebardFiory}, and in closely analogous systems like melting of two
dimensional crystals
\cite{HalperinNelson,Chui,Ceperley,BruunNelson} and dusty plasmas
\cite{Dusty:Plasma}.  More recently, features of the Kosterlitz-Thouless
transitions have been revealed in cold atoms experiments
with Bose-Einstein condensates in two dimensions
\cite{Dalibard,PolkovnikovKT}, including the observation of vortex
proliferation in two-dimensional Josephson arrays \cite{Cornell}.
Realizations of the clock model were also discussed in a variety of contexts, for example,
supersolid order of hard core bosons in triangular lattices
was predicted in Refs.~\onlinecite{Melko,Damle,Troyer}.
This type of model provides a good description for some frustrated
quantum magnet systems, where a two-step melting of the six-state
clock order via an intermediate, critical phase was recently
discussed \cite{Damle_new}.
We further note that the physics of the clock model and the existence of an intermediate phase has an analogue in lattice gauge theories in higher dimensions \cite{Elitzur,Ukawa}, where the critical phase corresponds to quantum electrodynamics with massless photons.

In the present paper this rich critical behavior is studied in the framework of a
concrete model of experimental relevance, namely, planar crystals of strongly interacting atoms, with a particular focus on realizations with trapped ions. Ion Coulomb crystals are an example
of ordered states of matter which emerge from the competition
between kinetic energy and repulsive forces in confined volumes
\cite{Dubin:RMP,Schiffer}. Their realization in atomic physics
laboratories  opens the possibility to gain insight into the dynamics
of diverse physical models, such as Wigner crystals
\cite{Wigner} of electrons on the surface of liquid Helium
\cite{Helium}, white dwarf cores, and neutron star outer
crusts \cite{vanHorn,Winget}. Thanks to the advances in trapping
and cooling \cite{VarennaBook}, different structures can be
realized by tuning the strength of the external confining
potential. The different structural order results from the
interplay between the repulsive interactions between the ions and
the trap geometry at low thermal energies, and has been
extensively analyzed in the classical regime by means of numerical
simulations \cite{dubin,Schiffer,Hasse} and of impressive
experimental realizations
\cite{Birkl92,Itano,2dion_exp,2dion_exp:1,Drewsen}.

\begin{figure}
\includegraphics[width=0.5\textwidth]{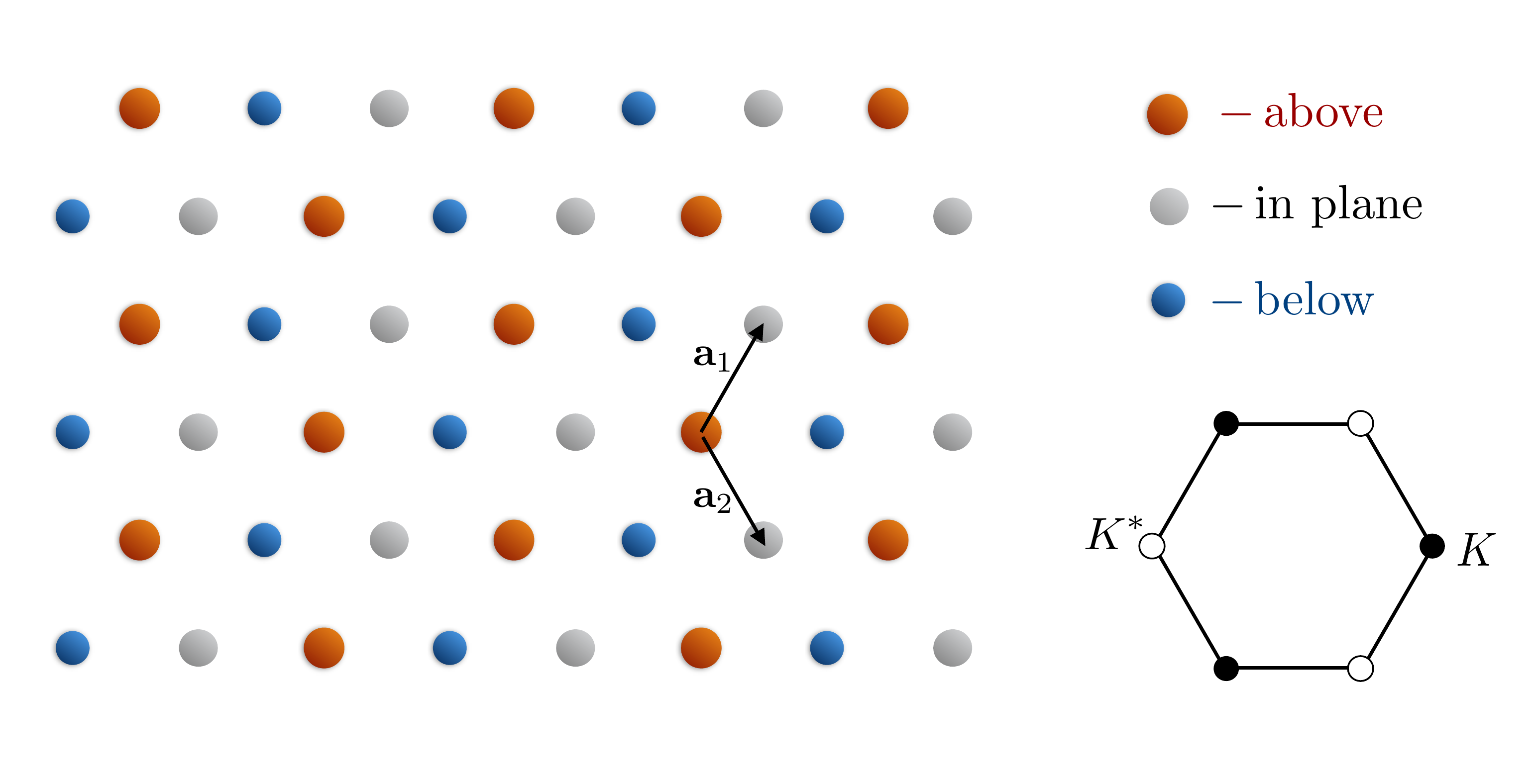} %
 \caption{(color online) Ions in a planar trap form a hexagonal lattice, which undergoes a mechanical instability as a function of the temperature and/or the transverse frequency. Left panel: Height pattern in the ordered phase. The hexagonal lattice is split into 3 sublattices, one of which stays in the $z=0$ plane (gray circles), one is raised above it (red), and one is lowered below it (blue).  There are 6 inequivalent configurations, corresponding to the $3!$ possible height assignments to the sublattices.  Right panel: The first Brillouin zone of the undistorted hexagonal lattice is shown, together with the wave vectors ${\bf K}$ and ${\bf K}^*=-{\bf K}$ corresponding to the height pattern. \label{fig:structure}}
\end{figure}

The manifestation of statistical mechanics and field-theoretical models in ion crystal systems has already been discussed in the context of one-dimensional ion chains.  For instance, the mechanical instability of a linear chain of ions to a zigzag structure can be classically mapped to a Landau model \cite{FCCM}, and the field theoretical model predicts a quantum phase transition of the Ising universality class \cite{SMF,SMFlong,varenna}. Careful numerical analysis based on DMRG-type of programs evaluated the quantum critical exponents with high precision and showed that the quantum critical region has a very small size because of the heavy ion mass \cite{DMRG:1,DMRG:2,varenna}.

The transverse instability of planar lattices of ions, by contrast, is largely unexplored, and offers the possibility of observing much richer physics. These two-dimensional structures are presently analyzed as platforms for quantum information processing \cite{Taylor,NIST_DD}. Early theory work by Dubin predicted a buckling
instability manifested as a continuous phase transition to a three-plane structure \cite{dubin}.   Similar results were obtained for a crystal of atoms interacting with repulsive dipolar interaction \cite{Demler}. Features hinting to the formation of three planes were measured in trapped ions experiments, even though these could not be revealed by Bragg scattering \cite{2dion_exp,2dion_exp:1}. The predicted structure is illustrated in Fig.~\ref{fig:structure}.

Intriguing properties are to be expected when fluctuations are taken into account.  The
buckling instability in Fig.~\ref{fig:structure} can be described by a complex-valued order parameter  ($\psi=|\psi |e^{i\theta}$), namely, the Fourier transform of the height configuration, evaluated at the wave vector ${\bf K}$ of the tripartite order.  Fluctuations in the phase $\theta$ of the order parameter are captured by the XY-model.   At yet lower temperatures, the phase $\theta$ is expected to be pinned to one of six favored discrete values $\theta_i$, leading to true long-range order and spontaneous breaking of $Z_6$ symmetry.  In this work we argue that the
planar-to-buckled instability is captured by the six-state clock
model \cite{Jose,Elitzur,Fradkin}.

As already noted above, the six-state clock model predicts the
existence of {\em two} transition lines at finite temperatures.
This behavior follows from a duality relation between the fully
ordered (low-$T$) and fully disordered (high-$T$) phases
\cite{Elitzur,Fradkin}, dictating two distinct critical
temperatures ($T_l$, $T_h$) at opposite sides of the self-dual
point. It implies the formation of an intermediate, critical phase
with power-law correlations in the finite range of temperatures
$T_l<T<T_{h}$. In our case, this physics is manifested as
follows: at fixed transverse trap frequency, there exists a
temperature $T_{KT}$ below which the planar crystal becomes
unstable, as shown in the phase diagram in Fig.~\ref{fig:pd}. Here, the ions' transverse displacements fluctuate
between three layers, and exhibit a buckled pattern with
quasi-long-range order. At a second temperature $T_6$, below
$T_{KT}$, true long-range order is established, leading to the
formation of three phase-locked layers, see Fig.
\ref{fig:structure}. The values of $T_{KT}$ and $T_6$, and hence
the size of the critical intermediate phase, depend on a
parameter $r$ tuned by the trap frequency, and is shown to vanish
at a single quantum critical point at $T=0$. By the general theory
of quantum phase transitions \cite{QPTbook} the model is
here in the universality class of the XY model in D=2+1
dimensions, up to dangerously irrelevant perturbations
\cite{Oshikawa}(for the formal definition see \onlinecite{danger}). Finally, for $T$ finite but sufficiently
small, we predict a crossover within the ordered phase from
classical to quantum behavior at a temperature $T_6^*<T_6$. This
signifies the emergence of an additional characteristic scale for
clock order, $\xi_6\sim 1/T_6^*$, which is intimately related to
the clock term being dangerously irrelevant \cite{Lou}.

As we show in this work, the realization of the quantum
clock model in cold ion systems provides unique access to features
that are typically inaccessible in standard condensed matter
systems (e.g. magnetic compounds).  In standard condensed matter
systems one usually measures global properties, such as
susceptibilities and conductivities.  By contrast, ionic traps and
cold atomic systems allow for the measurement of local properties through direct imaging,
and also offer continuous control of quantum parameters.  In
particular, a remarkable feature of the realization we propose in
trapped ions is that the different phases separated by the
buckling transitions are signaled by specific features of the
Bragg diffraction patterns, measurements which are typically
accessible in these systems \cite{Itano}. The line-shape of the
peaks, for example, provides a direct measurement of the
$T$-dependent exponent $\eta(T)$ characteristic of the
quasi-ordered Kosterlitz-Thouless phase.  In addition, we
show that by analyzing the shape of the transition lines,
the quantum critical properties
can be probed significantly far from the quantum critical point.
We further argue that these features can be observed in polarized
dipolar gases \cite{Lahaye,EugeneDipoles}, where the quantum
critical behavior may be manifested more prominently.

The outline of the paper is as follows: in Sec.~\ref{sec:op} we
introduce the order parameter field and derive the
field-theoretical model; technical details of the derivation are
included in Appendix A. In Sec \ref{sec:pd}, we analyze the phase
diagram.
In Sec.~\ref{sec:observables} (supplemented by Appendix B) we
derive the dependence of observable physical quantities in the
ordered phase on $T$ and the parameters of the model, particularly
analyzing the signature of the classical-quantum crossover
mentioned above. Our main conclusions are summarized in Sec.
\ref{sec:conclusions}.

\section{Derivation of a field-theoretical model}
\label{sec:op}

We consider a system consisting of repulsively interacting ions confined by a potential to move in the $x-y$ plane and its vicinity. The ions form a triangular lattice of spacing $a$  on the plane, as shown in Fig.~\ref{fig:structure}. We will assume that their only freedom is to move a small distance in the direction perpendicular to the lattice. This relies on the assumption, supported by numerical data \cite{dubin, Demler}, that the out-of-plane zigzag instability preempts a structural phase transition in the plane, {\em e.g.}, from a triangular to a square lattice. Throughout this paper we focus our discussion on ions, but these considerations can be extended to ultracold dipolar atoms or molecules, as we argue in Sec.~\ref{Dipoles}.

Note that throughout the paper (and unless otherwise stated) we work with units for which the Boltzmann and Planck constants are $k_B=\hbar=1$. All length scales are in units of the lattice constant $a$.

\subsection{Two dimensional ion crystals}

Ions of charge $Q$ and mass $M$ are confined in the $x-y$ plane by an anisotropic potential, whose strength is sufficiently large to overcome their mutual repulsion. We assume that the ion density $n$ on the plane is homogeneous, and denote by $U_c$ the potential confining the ions along the $z$ direction. The potential is harmonic and reads
\begin{equation}
\label{UCC}
U_c=\frac{C}{2}\sum_i z_i^2\,,
\end{equation}
where $z_j$ is the dimensionless displacement of ion $j$ from the plane (in units of the lattice spacing $a$), $C$ is a parameter with the dimensions of an energy: $C=M\omega_z^2a^2$, and $\omega_z$ is the harmonic-trap frequency. When the thermal energy is well below the Coulomb interaction energy the ions crystallize in a triangular lattice \cite{Dubin:RMP,Schiffer,Hasse}. In finite systems, this is observed at a sufficient distance from the edges \cite{2dion_exp,2dion_exp:1}. We denote by ${\bf r_j}=(x_j,y_j)$ the ions' positions on the plane, given in units of $a$. These are the solutions which minimize the in-plane potential energy $U_Q^{xy}$ given by
\begin{eqnarray}
U_Q^{xy}=\frac{\mathcal K}{2}\sum_{i \ne j}\frac{1}{\sqrt{(x_i-x_j)^2+(y_i-y_j)^2}}\label{UQQ:in plane}\,,
\end{eqnarray}
where ${\mathcal K}$ is the energy scale of the Coulomb  interaction,
\begin{equation}
\label{UCC1}
{\mathcal K}=\frac{Q^2}{4\pi \epsilon_0 a}\; ,
\end{equation}
with $\epsilon_0$ the vacuum permittivity.

In order to set up some notation, we consider a two dimensional (2D) hexagonal lattice with lattice constant $a=1$ and lattice vectors ${\bf a}_1$ and ${\bf a}_2$, given by
\begin{eqnarray}
{\bf a}_{1,2}&=&\left(1/2,\pm\sqrt{3}/2\right)\,.\end{eqnarray}
A general lattice vector can be written as
\begin{eqnarray}
{\bf r}_i=n_1{\bf a}_1+n_2{\bf a}_2\,,
\end{eqnarray}
with $n_{\{1,2\}}$ integers. The first Brillouin zone is a hexagon, with two inequivalent corners at the points $\pm {\bf K}$, where
\begin{eqnarray}
{\bf K}=(4\pi/3,0)\,.
\label{eq:K}
\end{eqnarray}
The other corner points are related to $\pm {\bf K}$ by reciprocal lattice vectors.

At fixed planar densities, the planar lattice is unstable against fluctuations in the $z$ directions, which tend to minimize the potential energy. This instability occurs at a critical value of the transverse potential, which depends on ${\mathcal K}$ and has been determined by means of a stability analysis based on minimization of the interaction energy in Ref.~\cite{dubin}.
For the purpose of our study, we consider the interparticle potential $U_Q$ resulting from the Coulomb repulsion after subtracting $U_Q^{xy}$:
\begin{eqnarray}
U_Q=\frac{\mathcal K}{2}\sum_{i \ne j}\frac{1}{\sqrt{(x_i-x_j)^2+(y_i-y_j)^2+(z_i-z_j)^2}}-U_Q^{xy}\,,\nonumber\\\label{UQQ}
\end{eqnarray}
and derive an effective field theoretical model assuming small displacements $|z_j|\ll 1$ from the plane at $z=0$.

\subsection{Planar instability and order parameter}

As the confining potential is reduced (or the planar density is increased), the system undergoes a
structural transition, in which some of the atoms leave the $z=0$ plane in order to reduce their mutual repulsion.
In the energetically favored configuration, the hexagonal lattice divides into a tripartite lattice. This is indicated by
direct minimization of the energy in numerical calculations \cite{dubin,Demler} and by analytical studies based on the Taylor expansion of the potential in Eq.~\eqref{UQQ} to  second order in the height $z$, which we report in App.~ \ref{UQT}. Particles on one sublattice rise, $z>0$, on another they are submerged, $z<0$, and on the third they remain level at $z=0$, as shown in Fig.~\ref{fig:structure}. 
Such a configuration is described by the height $z_i$ of the particle at the location ${\bf r}_i$, and can be succinctly written as
\begin{eqnarray}
z_i={\rm Re}\left[\psi e^{i {\bf K}\cdot {\bf r}_i}\right]\,,
\label{eq:heights}
\end{eqnarray}
where ${\bf K}$ is the vector defined in Eq.~\eqref{eq:K}, and $\psi=|\psi| e^{i\theta}$ is a complex number that acts as the
order parameter. Then, since
\begin{eqnarray}
{\bf K}\cdot {\bf r}_i=2\pi (n_1+n_2)/3\, ,
\label{eq:Kdotr}
\end{eqnarray}
the requirement that one of the three sublattices remains level at
$z=0$ implies that the phase of $\psi$ can only take one of six
values at the minimal energy configuration,
\begin{eqnarray}
\psi=|\psi| e^{i\theta}=|\psi| e^{i\pi(2n+1)/6}\,,
\label{eq:psi6}
\end{eqnarray}
where $n\in\{1,\ldots,6\}$.  This suggests that the structural
transition may be described in terms of a six-state clock model \cite{Jose} and is corroborated by a symmetry analysis, as follows.

The hexagonal lattice is symmetric under a number of transformations, including translations, reflections, and rotations.  These transformations act on both the lattice
positions ${\bf r}_i$ and on the heights $z_i$.  Equivalently, using
Eq.~\eqref{eq:heights}, these transformations can be thought of as
acting on the order parameter $\psi$.  For example, a reflection
about the $z=0$ plane, $R_z$, changes the sign of the heights
$z_i\to -z_i$ without affecting the positions ${\bf r}_i$.  This
is equivalent to flipping the sign of $\psi$,
\begin{eqnarray}
R_z:\psi\to -\psi\; .
\end{eqnarray}
On the other hand, a translation $T_{{\bf a}_1}$ by lattice vector ${\bf a}_1$  can be absorbed into a phase shift of $\psi$, since
\begin{eqnarray}
T_{{\bf a}_1}:\psi e^{i {\bf K}\cdot {\bf r}_i}&\to& \psi e^{i
{\bf K}\cdot ({\bf r}_i+{\bf a}_1)}\\&=&\left(\psi e^{2\pi
i/3}\right) e^{i {\bf K}\cdot {\bf r}_i}\, .
\end{eqnarray}
Hence,
\begin{eqnarray}
T_{{\bf a}_1}:\psi\to \psi e^{2\pi i/3}\; .
\end{eqnarray}
A translation by ${\bf a}_2$ acts the same way on $\psi$.  A third example is $R_x$, the reflection $x\to -x$.  Since ${\bf K}$ in Eq.~\eqref{eq:K} is parallel to $\hat{x}$, $R_x$ changes the sign of ${\bf K}\cdot {\bf r_i}$, that is
\begin{eqnarray}
R_x: {\rm Re}\left[\psi e^{i{\bf K}\cdot{\bf r}_i}\right]&\to& {\rm Re}\left[\psi e^{-i{\bf K}\cdot{\bf r}_i}\right]\\
&=&{\rm Re}\left[\psi^* e^{i{\bf K}\cdot{\bf r}_i}\right]\; .
\end{eqnarray}
Therefore,
\begin{eqnarray}
R_x:\psi\to \psi^*\,,
\end{eqnarray}
{\it i.e.}, $R_x$ acts as complex conjugation.

One can consider other symmetries of the hexagonal lattice, including rotations in the plane, the reflection $y\to -y$, and rotations by 180 degrees along an axis lying on the $z=0$ plane.  However, in all cases the action of these transformations will reduce to one of the cases studied above, namely:
\begin{eqnarray}
R_z: \psi&\to& -\psi\label{eq:minus}\,,\\
T_{{\bf a}_1}:\psi&\to&\psi e^{2\pi i/3} \label{eq:phaseshift}\,,\\
R_x: \psi&\to&\psi^*\label{eq:conj}\,,
\end{eqnarray}
or to combinations of these actions performed in sequence.  Hence, it is sufficient to consider these three basic actions.

\subsection{Mapping to a six-state clock model}

After coarse graining the lattice, one can write down a continuum Landau free energy for the order parameter $\psi$, which must be invariant under all the underlying symmetries of the lattice.  To sixth order in $\psi$ and $\psi^*$, and without derivatives, this restricts the allowed terms in the free energy to:
\begin{eqnarray}
|\psi|^2,\, |\psi|^4,\, |\psi|^6,\,
\end{eqnarray}
as well as
\begin{eqnarray}
\frac{1}{2}\left[\psi^6+(\psi^{*})^6\right] \,.
\end{eqnarray}
Note that the term $\frac{1}{2i}\left[\psi^6-(\psi^{*})^6\right]$ is forbidden since it is not invariant under $R_x$, Eq.~\eqref{eq:conj}.
In addition, the lowest order derivative term is
\begin{eqnarray}
|\nabla\psi|^2\,.
\end{eqnarray}
Thus, the Ginzburg-Landau (GL) free-energy density $f_{\rm GL}$ reads
\begin{eqnarray}
\frac{f_{\rm GL}}{\mathcal K}=
\frac{\gamma}{2} |\nabla\psi|^2+r|\psi|^2+u|\psi|^4+v |\psi|^6+ \frac{w}{2}\left[\psi^6+(\psi^{*})^6\right]\,,
\nonumber \\
\label{eq:GL}
\end{eqnarray}
which is here scaled by the Coulomb energy $\mathcal K$, Eq.~\eqref{UCC1}.
The  coefficients in Eq.~\eqref{eq:GL} can be obtained from the microscopic model
of Eqs.~\eqref{UQQ} and \eqref{UCC}  by expanding in small fluctuations starting
from the high symmetry, disordered phase. We detail this expansion in App.
\ref{UQT} for the case of various power-law repulsions including the Coulomb interaction.
For the specific case of the Coulomb interaction, the coefficient $\gamma$ is a number given by Eq.~\eqref{UgradCnum}.
Using (\ref{UC3}) and (\ref{UQ5}), the coefficient of the quadratic term reads
\begin{equation}
\label{rcoeff} r=\frac{1}{\sqrt{3}}\left(\frac{C}{2{\mathcal K}}-I_2\right)\,,
\end{equation}
where $I_n$ is a constant whose explicit form is given in Eq.~\eqref{In}.
For the coefficient of the quartic term, Eq.~(\ref{UQ54}) yields
\begin{equation}
\label{ucoeff}
u=\frac{3\sqrt{3}}{4} I_4.
\end{equation}
With the help of (\ref{UQ56}) we also find
\begin{equation}
\label{wcoeff}
w=\frac{5}{8\sqrt{3}} I_6
\end{equation}
and
\begin{equation}
\label{wcoeff}
v=-\frac{25}{4\sqrt{3}} I_6.
\end{equation}
Here,  $I_n$ and $\gamma$
are constants explicitly given in Appendix
\ref{UQT}. The sign of $w$
determines the values of the phase $\theta$ that minimize the free
energy. In particular, for $w>0$ this yields Eq.~\eqref{eq:psi6}.
We find that $v$ is negative, a situation that may result in a
first-order phase transition. In App.~\ref{UQT} it is checked that this
is not the case. This behavior can be taken into account by adding
a positive $8^{th}$ order term. Nevertheless, the correct physics can still be
captured by Eq.~(\ref{eq:GL}) by setting $v>0$, as assumed for simplicity
in the discussion of next Section.

When discussing quantum effects, we will need to add dynamics. The lowest order time-derivative term allowed by symmetry is
\begin{eqnarray}
|\partial_t\psi|^2
\end{eqnarray}
[Note that $i\psi^*\partial_t\psi$ is forbidden by Eq.~\eqref{eq:conj}].  Hence, the effective time-dependent Ginzburg-Landau Lagrangian is
\begin{eqnarray}
{\mathcal{L}}_{\rm GL}=\frac{\gamma}{2c^2}|\partial_t\psi|^2-f_{\rm GL}\,,
\label{eq:lagrangian}
\end{eqnarray}
where
\begin{eqnarray}
\label{eq:c}
c=\sqrt{\frac{\sqrt{3}\gamma{\mathcal K}}{2 M}}
\end{eqnarray}
is the speed of sound for transverse ($z$-polarized) phonons.

\subsection{Critical point of the mean-field model}

In order to connect with previous results, we remark that our theory naturally delivers the critical value of the ratio $C/{\mathcal K}$ of the mean field model, at which the planar crystal becomes unstable. This is found by setting $r=0$ in Eq.~\eqref{rcoeff} and yields the mean-field critical value
\begin{equation}
C_c^{\rm MF}=2{\mathcal K}I_2\approx 13.36 \mathcal K\,,
\end{equation}
that can be cast as a condition connecting the transverse trap frequency and the lattice constant:
\begin{equation}
\omega_z^{\rm MF}=\left(2I_2\frac{Q^2/(4\pi\epsilon_0)}{Ma^3}\right)^{1/2}\,.
\end{equation}
This value is consistent with that found in Ref.~\cite{dubin} (where it is reported in terms of the planar density $\sigma$).  We further point out that $C_c^{\rm MF}$ is an upper bound to the critical value. As we will see below, at finite temperature the planar instability is at a value $C_c(T)<C_c^{\rm MF}$, and decreases monotonically with $T$. At $T=0$, moreover, quantum fluctuations lower the value of the critical point from $C_c^{\rm MF}$ to $C_c(0)$ [so that $r_c<0$, see Eq.~(\ref{rcoeff})] by a quantity involving the ratio of kinetic and potential energies, as shown in Ref.~\onlinecite{podolsky}.

\subsection{Extension of the model to dipolar gases}
\label{Dipoles}

The mapping leading to the free-energy density in Eq.~\eqref{eq:GL} can be performed by using, instead of ions, dipolar atoms or molecules which are polarized by an external field perpendicular to the plane \cite{Lahaye,EugeneDipoles,RefDipoles}. In Appendix \ref{UDTALL} we derive the coefficients of the Landau free-energy density, corresponding to Eq.~\eqref{eq:GL}, as a function of the dipolar interaction. It must be kept in mind, however, that the dipolar interaction is not sufficiently long-range to warrant crystalline order in the plane, and the coupling with the planar modes can become relevant close to the mean-field critical point, changing the nature of the transition \cite{Larkin,Cartarius}. In principle, stability of the triangular lattice in the $x-y$ plane can be enforced by a pinning potential, e.g., by an optical lattice. For this realization, the free energy in Eq.~\eqref{eq:GL} describes the planar instability at small transverse displacements.

\section{Phase diagram}
\label{sec:pd}

\begin{figure}
\includegraphics[width=0.4\textwidth]{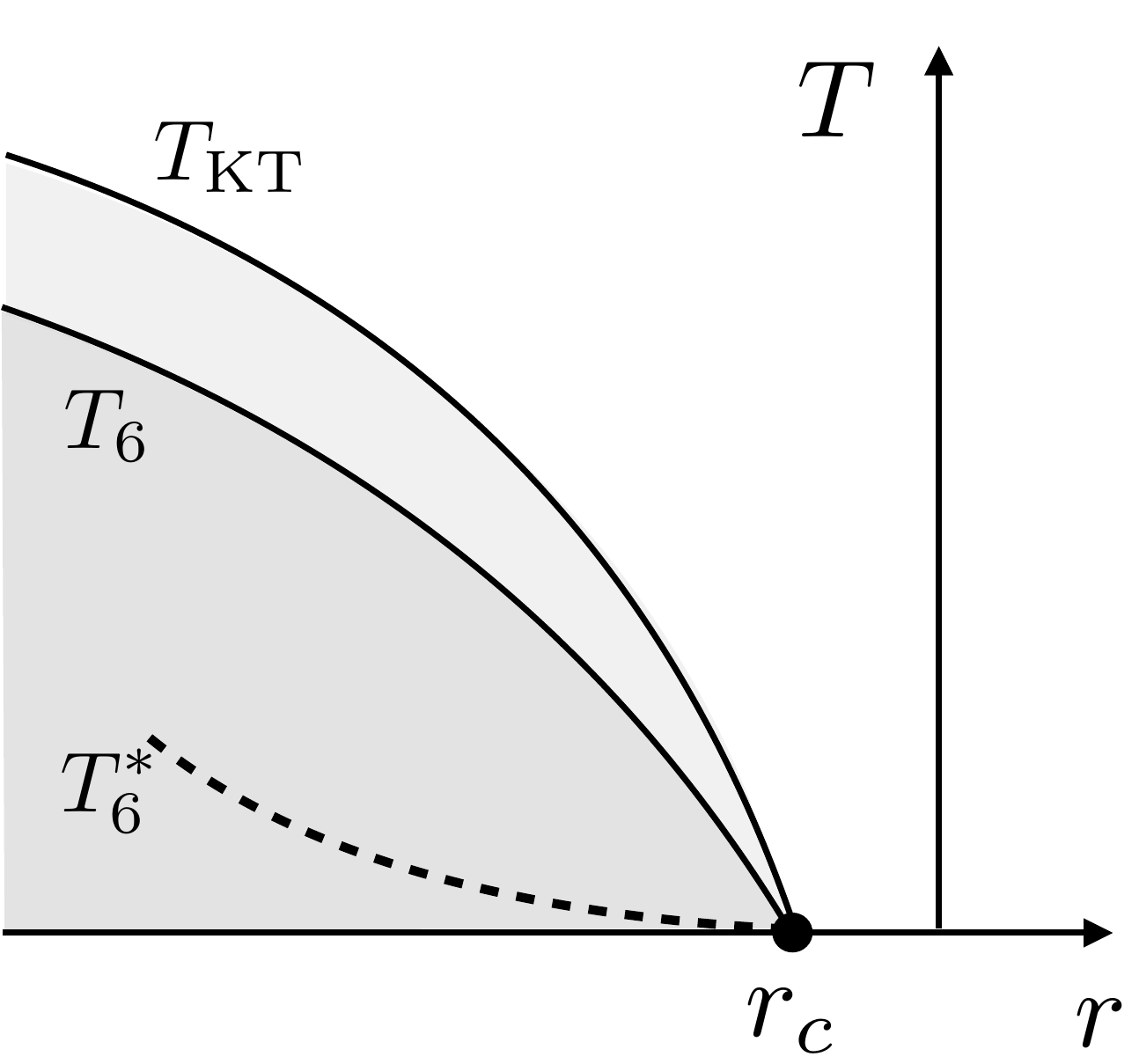}
 \caption{Phase diagram of the structural instability of a planar ion crystal in the parameter space determined by the temperature $T$ and the coefficient $r$, which is a function of the ratio ${\mathcal K}/C$. The external line at $T_{KT}$ indicates the transition from a planar structure to a critical phase, where the ions are distributed along three layers and exhibit quasi long-range order in the transverse direction. The second line at $T_{6}$ indicates the transition to a long-range ordered buckled phase, where the ions form three hexagonal lattices in the three layers which are phase locked as shown in Fig.~\ref{fig:structure}. The value $r_c<0$ indicates the critical point at $T=0$. The dashed line at $T_{6}^*$ separates the quantum from the classical critical behaviors. See text for further details.}
\label{fig:pd}
\end{figure}

The mapping of the model to the Landau-Ginzburg free-energy in Eq.~\eqref{eq:GL} allows us to determine the phase diagram of the system as a function of the temperature $T$ and of the ratio ${\mathcal K}/C$, as depicted in Fig.~\ref{fig:pd}. In ion Coulomb crystals, the first parameter is typically tuned by means of the laser which cools the ions, while the second ratio is controlled by changing the aspect ratio between the planar and the transverse trap frequencies.
We discuss below the phase diagram for the transition at finite temperature and for the quantum phase transition at $T=0$.

\subsection{Thermal transitions}
\label{sec:thermaltrans}

Let us first consider the phase diagram at finite temperature.  At mean-field level, assuming $u,v,w>0$ and $v>w$, the free energy in Eq.~\eqref{eq:GL} undergoes a second order phase transition when $r$ changes sign.  For $r>0$ the system is disordered, $\psi=0$, whereas for $r<0$ the system orders in one of
six minima of the free energy, $\psi=|\psi| e^{i\pi(2n+1)/6}$.

Beyond mean-field, in two spatial dimensions thermal fluctuations in the phase of the order parameter affect the phase diagram significantly. These fluctuations are captured by writing
$\psi=|\psi_0| e^{i\theta}$, where $|\psi_0|$ is the mean-field
value of the order parameter and $\theta$ is its phase.  Then,
Eq.~\eqref{eq:GL} yields
\begin{eqnarray}
f_{\rm GL}=\frac{\rho^0_s}{2}(\nabla \theta)^2+h_6 \cos (6\theta)\,,
\label{eq:clock}
\end{eqnarray}
where $\rho_s^0$ is the bare superfluid stiffness,
\begin{equation}
\label{rhos0}
\rho_s^0= \gamma \frac{{\mathcal K}}{2} |\psi_0|^2\,,
\end{equation}
and $$h_6=w |\psi_0|^6$$ is a term
that tends to pin $\theta$ in one of the six values given by
Eq.~\eqref{eq:psi6}. On the level of mean-field theory for the
model in Eq.~\eqref{eq:GL}, for $r<0$ the order parameter is different from zero and given by:
\begin{equation}
\label{psi0}
|\psi_0|^2=-\frac{r}{2u}=\frac{1}{9}\left(\frac{2I_2}{I_4}-\frac{C}{{\mathcal
K}I_4}\right)
\end{equation}
(a similar expression is found for dipolar interactions -- see App. A).
The resulting values of the parameters in Eq.~(\ref{eq:clock}) are
\begin{equation}
\label{rhos00} \rho_s^0= \gamma \frac{{\mathcal
K}}{2}\frac{1}{9}\left(\frac{2I_2}{I_4}-\frac{C}{{\mathcal
K}I_4}\right)\,,
\end{equation}
and
\begin{equation}
\label{h6} h_6=\frac{{5\mathcal K}}{8\sqrt{3}} I_6 \left[
\frac{1}{9}\left(\frac{2I_2}{I_4}-\frac{C}{{\mathcal
K}I_4}\right)\right]^3\; .
\end{equation}

Equation \eqref{eq:clock} defines the six-state clock model, which has been shown to exhibit two separate phase transitions \cite{Jose,Elitzur}.  Let us consider $r$ fixed, $r<0$. Starting from high temperatures (but still below the crystallization temperature), the system is in a disordered phase in the transverse direction, with exponentially decaying correlations. Then, as the temperature is reduced below a critical temperature, which we denote by $T_{\rm KT}$, vortices bind and the system undergoes a Kosterlitz-Thouless (KT) transition \cite{Berezinskii,KosterlitzThouless,ChaikinLubensky}.  The transition temperature is given by
\begin{equation}
\label{T:KT}
T_{KT}=\pi\rho_s/2\,,
\end{equation}
where the coefficient $\rho_s$ is the renormalized superfluid stiffness, that depends on $r$ and on temperature.  It is smaller than the bare superfluid stiffness $\rho_s^0$, but is typically of the same order of magnitude,  $\rho_s\lessapprox \rho_s^0$. At the transition, the clock coupling $h_6$ is irrelevant and an intermediate phase is reached in which the phase correlations are power-law in nature,
\begin{eqnarray}
\label{eq:powerlaw_corr}
\langle e^{i\theta({\bf r})} e^{-i\theta(0)}\rangle\sim |{\bf r}|^{-\eta}.
\end{eqnarray}
Here, $\eta$ is a continuous parameter which is monotonically decreasing with temperature, and can be expressed as
\begin{eqnarray}
\eta=T/(2\pi\rho_s)\,.
\label{eq:etarhos}
\end{eqnarray}
It attains the universal value $\eta=1/4$ at $T_{KT}$.

As the temperature is further reduced, when $\eta$ reaches the value $\eta=1/9$, {\em i.e.}, at the temperature $T_6<T_{\rm KT}$, with
\begin{equation}
\label{eq:t6}
T_6=\frac{2\pi\rho_s}{9}\,,
\end{equation}
the clock term becomes relevant
and the system undergoes an additional transition.  Below $T_6$, long-range $Z_6$ order is established, in which the phase of the order parameter is locked at one of the values in
Eq.~\eqref{eq:psi6}.

Let us now comment on the meaning of superfluid stiffness in the present context.  While the crystal of ions is clearly not superfluid, this terminology is borrowed from superconductivity and is broadly applied to XY models. In this case, it is a helicity modulus, which measures the free-energy increment associated with ``twisting" the direction of the order parameter, see Eq.~(\ref{eq:clock}).  Hence, it expresses the rigidity to deformations of the ordered pattern \cite{Fisher:1973}.

\subsection{Quantum critical point and its vicinity}
\label{sec:quantumcritical}

We next focus on the quantum phase transition at $T=0$. The temperature is a relevant variable at the quantum phase transition
and, therefore, the transition is in a different universality class from the thermal transitions at $T>0$.  Furthermore, in the
region surrounding the quantum critical point, the finite temperature phase diagram can be deduced from the scaling
properties of the quantum critical point \cite{SachdevRelax,FisherPfeuty}, as
described below.

The quantum phase transition is described by the Lagrangian in Eq.~\eqref{eq:lagrangian}.  Using the quantum to classical correspondence \cite{QPTbook} this can be mapped to a classical six-state clock model, Eq.~\eqref{eq:clock}, but now living in three dimensions.  The three-dimensional clock model
has been found to have a direct continuous transition between the long-range ordered $Z_6$ phase and the disordered phase.
The transition lies in the $XY$ universality class \cite{Hove,Blankschtein,Oshikawa,Lou}, with the addition of the clock term $h_6$ which is found to be dangerously irrelevant at the critical point \cite{Oshikawa,danger}. This implies that on both sides of the  $T=0$ transition, the quantum problem has a critical energy scale that vanishes at the transition as
\begin{eqnarray}
\Delta\sim|r-r_c|^{\nu},
\end{eqnarray}
where $\nu\approx 0.671$ is the correlation length critical
exponent of the $D=3$ $XY$ transition \cite{Campostrini}.  In addition,  when the
transition is approached from the ordered phase, the dangerously
irrelevant term $h_6$ introduces an additional critical energy
scale, whose effects will be discussed at the end of this section.

In the region of the phase diagram where $h_6$ is irrelevant ({\em
i.e.}, in the disordered and the critical phases above $T_6$ in
Fig.~\ref{fig:pd}),  $\Delta$ is the only critical energy scale.
Hence, in these regions, scaling near the quantum critical point
implies that  the two dimensional superfluid density behaves as
\cite{SachdevRelax}
\begin{eqnarray}
\rho_s (r,T)=T \Phi(\Delta/T)\,,
\end{eqnarray}
where $\Phi$ is a universal function of $\Delta/T$.  In particular, this implies that the contours of constant exponent $\eta$ in Eq.~\eqref{eq:etarhos} are also contours of constant $\Delta/T$.  Hence,
\begin{eqnarray}
T_{\rm KT}&=&A_{\rm KT} |r-r_c|^\nu \label{eq:TcScaling1}\,,\\
T_{6}&=&A_{6} |r-r_c|^\nu\label{eq:TcScaling2}\,,
\end{eqnarray}
where $A_{\rm KT}$ and $A_6$ are two non-universal constants. As
the stiffness $\rho_s$ is monotonic in $T$ at fixed $\Delta$ (it
is reduced by thermal fluctuations), and since $\eta$ is larger at
$T_{\rm KT}$ than at $T_6$, it follows that $A_{\rm KT}>A_6$.
Thus, we see that both transition temperatures go to zero at the
quantum critical point, with the same exponent but different
coefficients, and that the intermediate phase exists arbitrarily
close to the quantum critical point, although it becomes narrow.

We finally note that, according to Ref.~\onlinecite{Lou}, within
the ordered phase there is a length scale $\xi_6$ larger than the
$XY$ correlation length $\xi$.    In particular, near the quantum
critical point $\xi_6$ diverges faster, $\xi_6\sim \xi^{a_6}$
where $a_6\approx 9/4$.  For distances smaller than $\xi_6$, the
order parameter has effective $XY$ symmetry, and only for
distances beyond $\xi_6$ does the true long-range $Z_6$ order
become apparent \cite{Lou}.  By the quantum to classical
correspondence, this implies that in the $Z_6$ phase at $T=0$
there is a critical scale $\Delta_6$ which is softer than
$\Delta$, $\Delta_6\sim \Delta^{a_6}$.

At first glance, one may expect the scaling of $T_6$ near the
quantum critical point to be determined by $\Delta_6$ instead of
$\Delta$.  However, this is not the case since, when  $T_6$ is
approached from above, the relevance of the clock term  is
determined by the value of the superfluid stiffness alone.
This implies Eq.~(\ref{eq:TcScaling2}), as outlined above. One may
instead define a crossover temperature $T_6^\ast$, such that
\begin{equation}
T_6^\ast\equiv
\Delta_6\sim |r-r_c|^{a_6\, \nu}\,. \label{eq:tsixstar}
\end{equation}
Since $a_6>1$,  the temperature $T_6^*$ is
parametrically smaller than $T_6$ near the quantum critical point.
For finite temperatures in the range $T_6^\ast<T<T_6$ the
$Z_6$-ordering is characteristic of the classical $d=2$ model,
while for $T<T_6^\ast$ it is dominated by the quantum $XY$-model
in $D=d+1=3$.

\subsection{Practical considerations for experimental realizations on ion traps}

We now make a number of quantitative estimates of parameters
that may be useful in comparisons with experiments on ion traps.
First, it is useful to connect the temperatures appearing
in Sec.~\ref{sec:thermaltrans} with experimental values for
trapped ions. The features we predict will be observed below the
crystallization temperature $T_{\rm cryst}$, where $k_BT_{\rm
cryst}\simeq {\mathcal K}/\Gamma$ and $\Gamma$ is the plasma
parameter at which crystallization occurs, $\Gamma \approx 140$ in
two dimensions \cite{Hansen:1982} . Equations \eqref{T:KT} and
\eqref{rhos00} give $k_BT_{KT}\simeq  0.011\,{\mathcal
K}(I_2-C/2{\mathcal K})$. The temperature $T_{KT}$ can thus be a
fraction of the crystallization temperature depending on the
distance of the ratio  $C/{\mathcal K}$ from the mean field
critical value. For interparticle distances of the order of
15$\mu$m the crystallization temperature is in the milliKelvin
range \cite{Dubin:RMP}. For these parameters, sufficiently far
away from the mean field critical value, $T_{KT}$ can be of the
order of hundreds of $\mu$Kelvin. Comparing Eqs.~(\ref{T:KT}) and
(\ref{eq:t6}), the transition to long-range order is at a
temperature $T_6$ of the same order as $T_{KT}$.

These considerations suggest that the intermediate critical
phase could be observed using sub-Doppler cooling techniques
\cite{Eschner,Lechner}.  We remark that in general laser cooling can lead to different
effective temperatures for the planar and transverse modes of the Coulomb crystal \cite{2dion_exp:1}. During the experiment, thermalization between the vibrations is typically not observed, which
indicates that there is a sufficiently small coupling between axial and transverse vibrations, so that it can be neglected. This leads us to conjecture that corrections due to in-plane thermal fluctuations
can be neglected also sufficiently close to the transition, since in two dimensions the Coulomb interaction is sufficiently long-ranged to make the crystal incompressible
along the plane \cite{Larkin}.  This also corroborates the validity of our model, in which we assume that planar and transverse vibrations are consistently decoupled.

Next, we can estimate the size of quantum fluctuations in
Sec.~\ref{sec:quantumcritical} by computing the shift in quantum
critical point $r_c$ that they induce.  For this, we refer to Eq.
(2.12) in Ref.~\onlinecite{podolsky} which, in the notation of the present
paper becomes
\begin{eqnarray}
r_c=-\frac{3^{5/4}}{\pi\gamma^{1/2}}u\tilde{\hbar}
\end{eqnarray}
where $u$ is given by Eq. (2.26) while
\begin{eqnarray}
\tilde{\hbar}=\sqrt{\frac{\hbar^2}{2Ma^2 {\mathcal K} }}.
\end{eqnarray}
For $a=15\, \mu$m and single charge ions, this results in
\begin{eqnarray}
r_c=-3.8 \cdot 10^{-4}\frac{1}{\sqrt{N_A}}
\end{eqnarray}
where $N_A$ is the ion's atomic mass number.  For Beryllium, $N_A=9$, leading to
\begin{eqnarray}
r_c=-1.27\cdot 10^{-4}.\label{eq:rshift}
\end{eqnarray}
This corresponds to a fractional shift in the trap frequency
\begin{eqnarray}
\frac{\delta\omega_z}{\omega_z}=\frac{\sqrt{3}}{2 I_2} r_c\; .
\end{eqnarray}
For a trap of frequency 1 MHz, the shift is 17 Hz, and it could be measured when $\delta\omega_z T_h\ll 1$, with $T_h$ the heating time scale  of the ion trap \cite{2dion_exp:1,Brownnutt:2015}.

The shift in $r_c$ in Eq.~(\ref{eq:rshift}) can also be compared with the width of the buckled phase. Based on the stability analysis of Ref.~\onlinecite{dubin}, the three-layer phase is expected to exist over a range of planar densities $\sigma$, given by $\sigma a_0^2 \in [1.11,1.15]$, where $a_0$ is a length scale extracted from the trapping potential.   This somewhat narrow range is limited at high densities by a first order transition into a staggered square lattice structure, which occurs at  $\sigma=1.15/a_0^2$.  In terms of the parameter $r$, the lower density corresponds to $r=0$, where the mean-field onset of the three-layer occurs, whereas the higher density corresponds to
\begin{eqnarray}
r^{(1st\, \rm order)}=-\frac{\sqrt{3} I_2}{2} \frac{\delta \sigma}{\sigma}=-0.21\, ,\label{rfirstorder}
\end{eqnarray}
where $\delta \sigma/\sigma=(1.15-1.11)/1.11$ is the relative width of the phase.  Hence, the fluctuation-induced shift in Eq.~(\ref{eq:rshift}) is very small relative to the full width of the phase.

Another practical issue to address is the effect of the finite system size. For harmonic confinement, the triangular crystalline structure can be observed with 1000 ions at the center of the plane \cite{Schiffer}.   Present experiments can realize planes with about 1000 ions, corresponding to a linear dimension of $L\approx 30$ ions.  Deep in the ordered and disordered phases, when the correlation length $\xi$ is well below $L$, the finite size of the system is not important.  By contrast, as the critical phase is approached within either of these phases, $\xi$ diverges and the finite system size must be taken into account.  In particular, when $\xi$ is of order $L$, the system will {\em appear} to be critical.  This places a practical limitation in mapping out the phase transition lines.  However, by studying the dependence on system size, it is possible to locate these transitions with high accuracy using relatively modest system sizes.  For example, in Ref.~\onlinecite{WeberMinnhagen}, the Kosterlitz-Thouless temperature was determined with accuracy of three significant digits by simulating systems of size $12\times 12$ and smaller.

On the other hand, within the critical phase itself, the finite system size does limit the ability to accurately measure critical exponents, since $L=30$ gives access to only a relatively short segment of a power law.  A rough rule of thumb is that one can extract one significant digit for each decade of scaling, so that here we may expect to obtain the first digit of the critical exponents.
Finally, the observation of the scaling form for $T_6^*$, given in Eq.~(\ref{eq:tsixstar}), will require larger system sizes than 1000 atoms. Close to the quantum critical point, in fact, this crossover occurs at a long correlation length which is associated with the dangerously irrelevant clock term.  Far  from the critical point,  $T_6^*$ might be extracted from the temperature dependence of the clock order parameter, as discussed below.

\section{Measurable signatures of the phase diagram}
\label{sec:observables}

We now focus on physical quantities that are in principle accessible to experimental probes. We first summarize our findings.
The mechanical instability of the two-dimensional Coulomb crystal gives rise to a transition to order in the transverse direction, which occurs in two stages:  for temperatures $T$ between $T_6$ and $T_{KT}$ there is a critical phase with quasi-long-range order \cite{Jose}. For $T<T_6$, in the ordered phase, the system spontaneously selects one of $6$ equivalent patterns, {\em i.e.} breaks a $Z_6$-symmetry. This structural phase transition is therefore a realization of the ordering in a six-state clock model, manifested here by splitting of the single layer of ions into three layers, at $z=0,\pm h$. Hence the layer separation $h$ serves as a measurable probe of the order parameter. In addition, the resulting enlargement of the unit cell in the ordered pattern Fig.~\ref{fig:structure} is expected to induce new Bragg peaks in a diffraction experiment. Their height and line-shape as functions of $T$ and the system parameters can provide a further test of the predictions of the theoretical model. In this Section we also propose a possible observable for the classical-quantum crossover within the ordered phase at a characteristic temperature $T_6^*$. For temperatures below $T_6^*$, the layer separation $h(T)$ is expected to saturate to a constant value.  This criterion will serve as an operational definition of $T_6^*$.

The discussion of this section is centered on determining the signatures of the phase diagram by direct measurement of the crystal planes and by a Bragg diffraction experiment.  Bragg scattering could be performed at the asymptotics of laser cooling, when the ions have reached the stationary state and the emitted photons carry the information about the crystal structure and excitations\cite{Itano,Dantan}. In addition, it is worth mentioning that the transverse shift of the crystal planes could  also be measured by mapping the crystal structure into the electronic degrees of freedom of the ions \cite{DeChiara:2008,Retzker:2008,2dion_exp:1,Demler-Bloch}. These methods can in principle provide precise information of the transverse structure and fluctuations. Moreover, thanks to the advances in high precision spectroscopy, such measurements can provide the precision required to access the quantum regime.

\subsection{Temperature dependence of the layer separation}

Assuming that direct imaging of
the particles is possible, the ordering established in the dark
shaded area of Fig. 2 would be manifested as splitting of the
plane to three layers separated by a distance
\begin{equation}
h={\rm max}_i\left(|\langle z_i \rangle |\right)\,,
\end{equation}
where $z_i$ is given by Eq.~(\ref{eq:heights}) with $i$
on the $+$ or $-$ sublattice of the broken-symmetry pattern (Fig.
1), and $\langle ... \rangle$ denotes the thermal expectation value.
Expressing the phase of the order parameter field $\psi$ as
$\theta=\theta_0+\delta\theta$, where $\theta_0$ is one of the
clock states $\frac{\pi}{6}(2n+1)$, one obtains
\begin{eqnarray}
h=|\psi_0|\cos(\pi/6)\langle \cos(\delta\theta) \rangle\; .
\label{eq:h_def}
\end{eqnarray}
where $\psi_0$ is given in Eq.~(\ref{psi0}).
While in the intermediate critical phase the ions exhibit thermal fluctuations between the three layers, for $r<r_c$ and $T<T_6$ the expectation value $\langle \cos(\delta\theta) \rangle\ne 0$ and consequently $h$ is finite in the thermodynamic limit. It can be evaluated using an effective Gaussian theory for the phase fluctuations $\delta\theta$ (see Appendix \ref{app:var} for details). This yields
\begin{eqnarray}
h(T)|_{r<r_c,T<T_6}=h_0 \left[2\sinh\left(\frac{m(T)}{2T}\right)\right]^{\frac{\eta(T)}{2}}\,;
\label{eq:hvsT}
\end{eqnarray}
here $m(T)$ is a temperature dependent coefficient, whose explicit form is given in Appendix \ref{app:var}, and
\begin{eqnarray}
h_0=|\psi_0|\cos(\pi/6)e^{-\frac{c \Lambda}{8\pi\rho_s}}\label{eq:h0}
\end{eqnarray}
is a non-universal amplitude which depends on the high momentum cutoff $\Lambda\approx \pi/a$. In the quantum regime, {\em i.e.}, for $T\ll T^*_6$ (below the dashed line in Fig.~\ref{fig:pd}), Eq. \eqref{eq:hvsT} can be approximated by the temperature-independent value
\begin{eqnarray}
h(T)|_{T\ll T^*_6}=h_0 e^{\frac{m_0}{8\pi\rho_s}}\,,
\label{eq:hvsTQ}
\end{eqnarray}
where $m_0$ is given in Eq.~(\ref{m0def}) of App.~\ref{app:var}. In fact, the saturation itself of $h(T)$ occurs at $T\approx m_0$, which according to our definition, allows us to identify
\begin{eqnarray}
T_6^*\approx m_0\,.
\end{eqnarray}
The saturation value, Eq.~\ref{eq:hvsTQ}, becomes invalid close to the quantum critical point ($r\rightarrow r_c$, $T=0$) where
\begin{eqnarray}
h|_{r\to r_c,T=0}\sim |r-r_c|^\beta\,,
\end{eqnarray}
with $\beta\approx 0.348$ the $XY$ critical exponent \cite{Lou,Campostrini} at $D=3$. In the classical regime ($T^*_6\ll T<T_{6}$)
\begin{eqnarray}
h(T)|_{T^*_6\ll T<T_{6}}&=&h_0\left(\frac{m(T)}{T}\right)^{\frac{\eta(T)}{2}}\\
&\approx &h_0
\left(\frac{m_0}{T}\right)^{T\over 18(T_6-T)}
\label{eq:hvsTC}\,,
\end{eqnarray}
where in the final expression we  employed Eq.~(\ref{m_vs_T_class})
of App.~\ref{app:var} and also Eqs.~(\ref{eq:etarhos}) and (\ref{eq:t6}).

\begin{figure}
\includegraphics[width=0.5\textwidth]{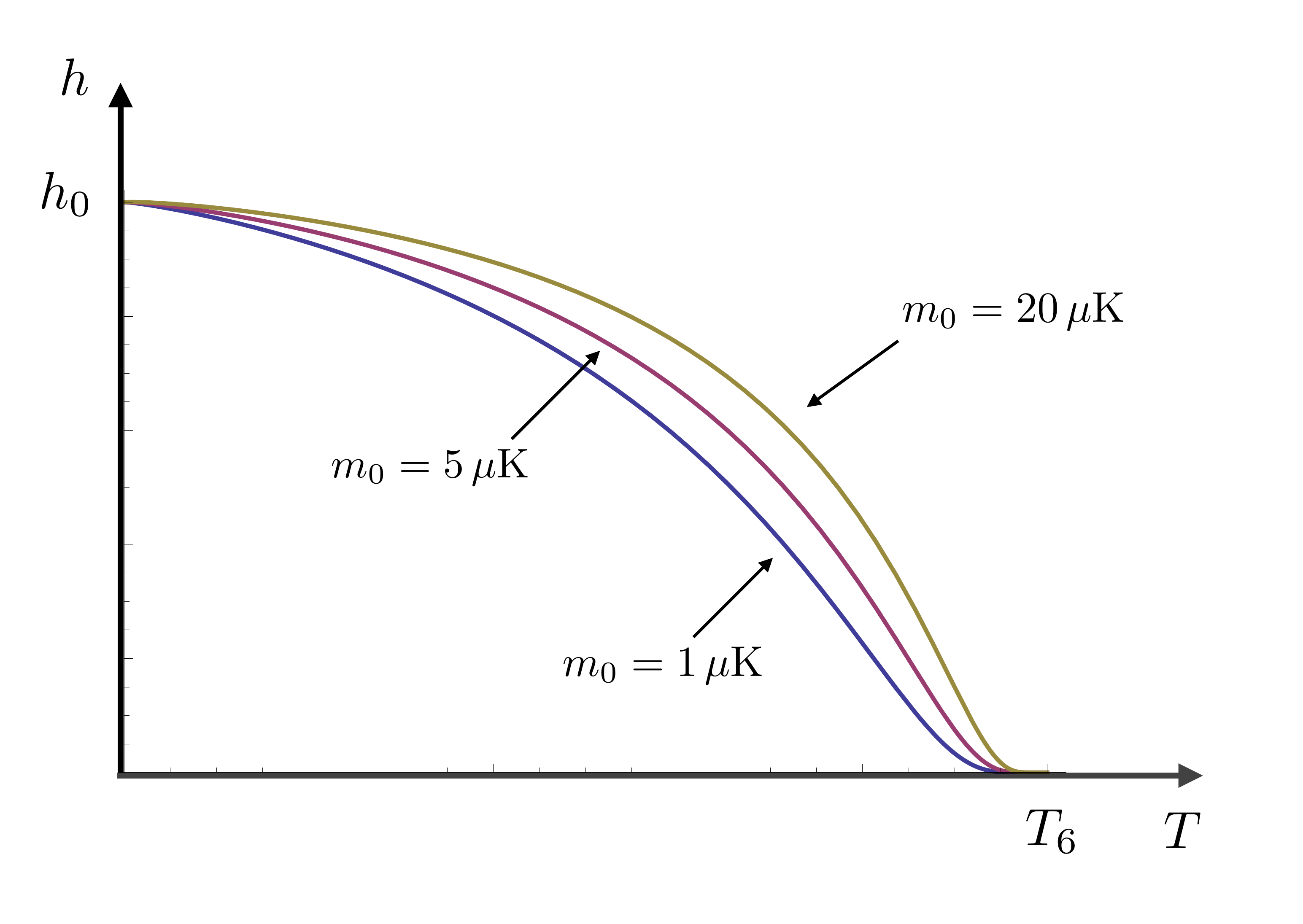} %
 \caption{ (color online) Interlayer separation $h(T)$ as a function of temperature, based on Eq.~(\ref{eq:hvsTC}). The zero temperature value, $h_0$, is of order one micron, as shown in Eq.~(\ref{eq:h0}). Note that the functional form of $h(T)$ is sensitive to the value of $m_0$, even at temperatures well above $m_0$ itself.  Throughout, the onset temperature for clock order is taken to be $T_6=10$ mK. \label{fig:hT}}
\end{figure}

In oder to illustrate the temperature dependence of $h(T)$, we consider beryllium ions with lattice constant $a=15\,\mu$m.  Then, combining Eqs.~(\ref{rcoeff}), (\ref{rhos00}), (\ref{h6}) and (\ref{m0def}), we obtain
\begin{eqnarray}
h_0&=&(4.3\, \mu {\rm m})\cdot \sqrt{|r|}\,,\\
m_0&=&(1.3\times 10^{-5}\,{\rm Kelvin})  \cdot |r|\,.
\end{eqnarray}
These expressions are mean-field in nature and do not take fluctuations into account, but they do give rough estimates for the values of the parameters.  In particular, since $r$ is bounded by Eq.~(\ref{rfirstorder}), $|r|<0.21$, $m_0$ is in the $\mu$K range (by comparison, for similar parameters, $\rho_s$ is of order a few mK).  This value is small compared to typical experimental temperatures.  Hence, Eq.~(\ref{eq:hvsTC}) provides a good approximation to the order parameter $h(T)$.  Figure \ref{fig:hT} shows the temperature dependence of $h(T)$  for a number of values of $m_0$.  Note that although $m_0$ of this magnitude is too small to be accessed directly in experiments,  its value affects the temperature dependence of $h(T)$ at $T\gg m_0$.  Hence, it may be possible to extract an estimate for $m_0$ (and therefore of $T_6^*$) from a fit to $h(T)$, using Eq.~(\ref{eq:hvsTC}).

\subsection{Diffraction pattern}

Another measurable quantity is the diffraction pattern $I_{{\bf k}_{3d}}$ at wave-vector ${\bf k}_{3d}=({\bf k},k_z)$, where ${\bf k}$ is a vector in the $x-y$ plane. The diffraction pattern probes the structure factor $S_{{\bf k}_{3d}}$, that is the Fourier transform of the density operator, via the quantity
\begin{eqnarray}
I_{{\bf k}_{3d}}= \langle |S_{{\bf k}_{3d}}|^2 \rangle\; .
\label{eq:I2S_k}
\end{eqnarray}
For $k_zh\ll 1$, this expression can be cast in the form
\begin{eqnarray}
I_{{\bf k}_{3d}}=I_{({\bf k},k_z=0)}+F({\bf k},k_z)\,,
\end{eqnarray}
with the leading $k_z$-dependent contribution
\begin{eqnarray}
F({\bf k},k_z)&= & \frac{k_z^2|\psi_0|^2N}{4}\sum_{j}\left(e^{i({\bf k}+{\bf K})\cdot{\bf r_j}}+e^{i({\bf k}-{\bf K})\cdot{\bf r_j}}\right)C({\bf r_j})\nonumber\\
C({\bf r})&\equiv & \langle e^{i(\delta\theta({\bf r})-\delta\theta(0))} \rangle
\label{eq:F_k2C(r)}
\end{eqnarray}
(details are given in Appendix \ref{App:C}).
This contribution exhibits peaks at new wave vectors ${\bf k}_0=\pm {\bf K}+{\bf
G}$, where ${\bf K}$ is given by Eq.~(\ref{eq:K}) and ${\bf G}$ is a reciprocal lattice vector, reflecting
the enlarged unit cell. Defining the small deviation ${\bf q}={\bf
k}-{\bf k}_0$, we get
\begin{eqnarray}
F({\bf q},k_z)\approx \frac{k_z^2|\psi_0|^2N}{4}\int d^2r\,e^{i{\bf
q}\cdot{\bf r}}C({\bf r})\; . \label{eq:F_q2C(r):0}
\end{eqnarray}
Within the Gaussian effective theory (Eq.~(\ref{eq:gaussian}) in Appendix \ref{App:C}), the asymptotic form of $C({\bf r})$ for large ${\bf r}$ is
\begin{eqnarray}
C({\bf r})&\approx & \langle \cos(\delta\theta)
\rangle^2\left(1+\langle \delta\theta({\bf r})\delta\theta(0)
\rangle\right)\nonumber\\&=&{\rm Const.}+\delta C({\bf r}) \label{eq:gaussian_main}\,,
\end{eqnarray}
where in the entire ordered phase ($T<T_6$, $r<r_c$) the correlation function $\delta C({\bf r})$ is
exponentially decaying over a length $c/m$ [see Eq.~(\ref{eq:theta_corr_T})]. Consequently, the
line-shape of the new diffraction peaks [$F({\bf q},k_z)$ defined
in Eq.~(\ref{eq:F_q2C(r):0})] will be given by $\delta({\bf
q})+F_{\rm fluc}({\bf q})$ where the fluctuations contribution
$F_{\rm fluc}({\bf q})$ has the form of a Lorentzian with a
characteristic width $\sim m(T)/c$.

We finally focus on the critical phase established in the higher $T$ regime $T_6<T<T_{KT}$ (the light-shaded region in Fig.~\ref{fig:pd}), where quasi-long-range order is anticipated. The hallmark of such phase is a critical behavior of the correlation functions $C({\bf r})$, which should manifest the power-law decay of Eq.~(\ref{eq:powerlaw_corr}) with the $T$-dependent exponent $\eta$. As $T$ is reduced from the upper critical temperature $T_{KT}$, this exponent varies continuously from $\eta=1/4$ at $T_{KT}$ to $\eta=1/9$ at $T_6$. We expect that this behavior will be reflected by the appearance of smeared Bragg peaks at the new wave-vectors ${\bf k}_0$. Their line-shape $F({\bf q},k_z)$ should exhibit the power-law dependence \cite{HalperinNelson}
\begin{eqnarray}
\label{Fq_eta}
F({\bf q},k_z)\sim |{\bf q}|^{\eta-2}\; .
\end{eqnarray}
Hence, in principle a diffraction measurement can serve as direct probe of the exponent $\eta(T)$.

\section{Conclusions}
\label{sec:conclusions}

In conclusion, we have argued that the mechanical instability of a planar crystal of trapped ions can be mapped to the six-state clock model, and thus manifests a two-step buckling transition. On the basis of symmetry consideration we derived the field theoretical model and determined the coefficients starting from the microscopic Hamiltonian. The phase diagram is characterized by several phases, depicted in Fig.~\ref{fig:pd} as a function of the temperature $T$ and of the external trap frequency, here parametrized by the coefficient $r$.  It exhibits an intermediate, critical phase between disorder and long-range order in the buckled pattern of transverse displacements, within a finite range of parameters. This intermediate phase shrinks to zero as $T\to 0$, where a unique critical point is expected.  Finally, we argued that within the ordered phase, a crossover from a classical to a quantum behavior occurs at a yet lower temperature $T_6^*(r)<T_6$. This signifies the emergence of an additional characteristic scale for clock order, $\xi_6\sim 1/T_6^*$, which is intimately related to the clock term being dangerously irrelevant.

These behaviors are manifested in the functional dependence of the interlayer distance $h$ and of the Bragg signal, accessible in experiments with trapped ions, on the trap frequency and on the temperature of the crystal. In particular, in a pancake-shaped trap of the type used in Ref.~\onlinecite{2dion_exp}, a gradual change of particle density towards the center of the $x-y$ plane can yield a ``wedding cake'' structure \cite{weddingCake}, and would enable the probing of $h$ {\it vs.} the tuning parameter $r$ as it gradually varies with the distance from the center.

We finally note that according to our estimate of the parameters for ion crystal systems, the thermal transition lines $T_{KT}(r)$ and $T_6(r)$ are expected to be observable with present trapping and cooling techniques.  Direct probing of the quantum regime (by cooling to temperatures of order $T^*_6$) is more challenging.  However, information on quantum effects can be drawn from measurements done at accessible, higher temperatures ($T\gg T_6^*$).  For example, by fitting the curve $h$ {\em vs.}~$T$ to the theory, the energy scale $T_6^*$ can be extracted. In addition, by analyzing the critical temperatures $T_{KT}$, $T_6$ as a function of $r$, it may be possible to verify the critical exponent for the quantum transition [see Eqs.~(\ref{eq:TcScaling1}) and (\ref{eq:TcScaling2})], even far from the quantum critical point.
Furthermore, the enhanced quantum fluctuations in systems of ultracold dipolar gases, such as in the setup of Ref.~\onlinecite{RefDipoles}, provide an independent realization in which quantum effects may be observed and measured more directly.

\acknowledgements
We are grateful to Joe Britton, John Bollinger, Alexey Gorshkov, Shlomi Kotler, Mikhail Lukin, Yoav Sagi, and especially to Eugene Demler, for illuminating discussions. This work was partially supported by the German Research Foundation (DFG), the Israel Science Foundation (ISF)
grant numbers 1839/13, 231/14, and 1028/12, the Joint UGS-ISF Research Grant Program under grant number 1903/14, the US-Israel Binational Science Foundation (BSF) grant number 2010132, and by the Shlomo Kaplansky academic chair.  ES, GM, and SF thank the Kavli Institute for Theoretical Physics in Santa Barbara for its hospitality, where this research was
supported in part by the National Science Foundation (NSF) under grant number
NSF PHY11-25915.  DP and ES thank the Aspen Center for Physics, where part of this work was done, with support of NSF grant number PHY-1066293. GM thanks the Ion Storage Group at NIST, Boulder, for hospitality and support during completion of this work.

\appendix

\section{Derivation of the Ginzburg-Landu (GL) free energy}
\label{UQT}

Long-range order results in the increase of the unit cell, therefore we turn first to find the wave vectors of the long-range order. Since the interaction is the strongest for nearest neighbors we assume the nature of the long-range order is determined by the nearest neighbors, while further neighbors determine the numerical values of parameters. This assumption is tested numerically.

\subsection{The order determined by nearest neighbor interactions}
\label{UQNN1}

The potential energy for nearest-neighbor interactions can be written after truncating the Coulomb sum, and takes the form
\begin{equation}
\label{VV}
V=\frac{C}{2}\sum_iz_i^2-\frac{{\mathcal K}}{4}\sum_{\langle i,j\rangle} (z_i-z_j)^2\,,
\end{equation}
where $\langle i,j\rangle$ denotes the sum over nearest neighbors. Since the long-range order is characterized by a wave vector, it is convenient to introduce the Fourier Expansion of the transverse displacement,
\begin{equation}
\label{VF1}
z_i=\sum_{\bf k} \tilde{z}_{\bf k}e^{i {\bf k}\cdot{
\bf r}_i}\,,
\end{equation}
where ${\bf r}_i$ is a point on the two dimensional lattice end $z_i$ is the distance of ion at this point from the plane.
In Fourier space the first term in Eq.~(\ref{VV}) reads
\begin{equation}
\label{VF2}
V_1=\frac{C}{2}\sum_iz_i^2=\frac{C}{2} \sum_{\bf k}|\tilde{z}_{\bf k}|^2\,,
\end{equation}
where we have used that $ {z}_{\bf -k}={z}_{\bf k}^*$, being $z_i$ real.
The second term on the right-hand side of Eq. \eqref{VV} is given by
\begin{equation}
\label{VF3}
V_2={\mathcal K} \sum_{\bf k}|\tilde{z}_{\bf k}|^2F({\bf k})\,,
\end{equation}
with
\begin{equation}
F({\bf k})=\sum_{i=1}^6 \sin^2\left(\frac{1}{2}{\bf k}\cdot\delta {\bf r}_i\right)\,.
\end{equation}
Here, $\delta {\bf r}_i $ are the vectors that join some site of the planar lattice to the corresponding six nearest neighbors, they are combinations of ${\bf a}_{1,2}$ with integer coefficients $\pm 1,0$. Using their explicit form,
\begin{equation}
F({\bf k})=3-\cos k_x-2\cos \frac{1}{2}k_x \cos \frac{\sqrt{3}}{2}k_y\,.
\end{equation}
Finally, in Fourier space potential \eqref{VV} takes the form
\begin{equation}
\label{VVF}
V=V_1-V_2=\sum_{\bf k}|\tilde{z}_{\bf k}|^2\left(\frac{C}{2}-{\mathcal K} F({\bf k})\right)\,.
\end{equation}
For sufficiently large confinement $C$ the ions are confined to the $x,y$ pane. This is the high-symmetry disordered phase. According to Landau theory, long-range order is expected for a vector ${\bf k}=(k_x,k_y)$ for which the sign of the corresponding term in the sum (\ref{VVF}) first changes sign as $C$ is increased. This happens for the value of ${\bf k}$
that maximizes $F({\bf k})$. A straightforward calculation shows that the appropriate wave vectors are the corners of the Brillouin zone $\pm{\bf K}$ where ${\bf K}$ is given by Eq.~(\ref{eq:K}), namely
\begin{equation}
\label{kmax}
{\bf K}=\left(\frac{4\pi}{3},0\right),
\end{equation}
as well as for other vectors related to them by the reciprocal lattice vectors. In particular, for ${\bf K}$ in Eq.~(\ref{kmax}), $F({\bf K})=\frac{9}{2}$, and  in the vicinity of the critical point the energy related to the long-range order reads
\begin{equation}
\label{UK}
U_{{\bf K}}=\left(\frac{C}{2}-\frac{9}{2}{\mathcal K}\right)|z_{{\bf K}}|^2\,.
\end{equation}
The contribution of the nearest neighbors to the gradient term in the LG free-energy density is
\begin{equation}
\label{vfprime}
V'_f=\frac{3}{4}{\mathcal K}|\delta {\bf k}|^2|\tilde{z}_{\bf K}|^2\,,
\end{equation}
where $\delta {\bf k}$ is the deviation from $K$, or
\begin{equation}
\label{vfluc}
V'_f=\frac{3}{8}{\mathcal K}|\nabla\psi|^2\,,
\end{equation}
and which thus delivers the value of the constant $\gamma$ in Eq. \eqref{eq:GL} for nearest-neighbour interactions.
In the following, the contribution of farther neighbors will be calculated.

\subsection{Parameters of the GL free energy for the Coulomb potential}
\label{UQAll}

We now derive the parameters of the Ginzburg-Landau free energy, Eq. \eqref{eq:GL}, taking the full Coulomb interaction of Eq. \eqref{UQQ}.
Using the order parameter (\ref{eq:heights}) the distance from the $x,y$ plane takes the form
\begin{equation}
\label{order-parameter}
z_i=|\psi|\cos({\bf K}\cdot{\bf r}_i+\theta)\,,
\end{equation}
where
\begin{equation}
\label{coordinate}
{\bf r}_i \equiv {\bf r}(n_{1,i},n_{2,i})=\left(\frac{1}{2}(n_{1,i}+n_{2,i}),\frac{\sqrt{3}}{2}(n_{1,i}-n_{2,i})\right)\,,
\end{equation}
with $n_{1,i},n_{2,i}=0,\pm1,\pm2,\ldots$, and ${\bf K}$ is given in Eq. \eqref{eq:K}. Substituting Eq. \eqref{order-parameter} into Eq. \eqref{UCC}, the contribution of the harmonic potential to the GL free energy is written as
\begin{equation}
\label{UC3}
U_c=N\frac{C}{4}|\psi|^2=\Omega \frac{C}{2\sqrt{3}} |\psi|^2\,,
\end{equation}
where $N$ is the number of ions and
\begin{eqnarray}
\Omega=\frac{\sqrt{3}}{2} N
\end{eqnarray}
is the area of the sample (in units of $a^2$). In order to obtain the GL free energy  of Eq. \eqref{eq:GL} we expand the Coulomb energy $U_Q$, Eq.~(\ref{UQQ}), in powers of $z_i$. Being the plane an equilibrium configuration, the first order term vanishes. The second-order term reads:
\begin{equation}
\label{UQ3}
U_{2,Q}= -\frac{{\mathcal K}}{4} \sum_{i,j} \frac{(z_i-z_{j})^2}{|{\bf r}_i-{\bf r}_j|^3}\,,
\end{equation}
and in terms of the order parameter $\psi$ of Eq, (\ref{eq:heights}) can be recast in the form:
\begin{eqnarray}
\label{UQ4}
(z_i-z_j)^2&=&4|\psi|^2\sin^2\left(\frac{{\bf K}\cdot ({\bf r}_j-{\bf r}_i)}{2}\right) \\
&\,&\,\times\sin^2\left(\frac{{\bf K}\cdot ({\bf r}_j-{\bf r}_i)}{2}+{\bf K}\cdot{\bf r}_i+\theta\right)\nonumber\,.
\end{eqnarray}
Performing the sum in Eq.~(\ref{UQ3}) over $i$ for fixed ${\bf (r}_j-{\bf r}_i)$ (that can be considered a lattice vector of the form (\ref{coordinate})), we find
\begin{equation}
\label{UQ5}
U_{2,Q}=-{\Omega{\mathcal K} \over \sqrt{3}}|\psi|^2I_2\,,
\end{equation}
where
\begin{equation}
\label{In}
I_n=\sum_{n_1,n_2} \frac{\sin^n\left[\frac{\pi}{3}(n_1+n_2)\right]}{|{\bf r}_{n_1,n_2}|^{n+1}}\,,
\end{equation}
with
\begin{equation}
\label{absr}
|{\bf r}_{n_1,n_2}|=\sqrt{\frac{1}{4}(n_1+n_2)^2+\frac{3}{4}(n_1-n_2)^2}.
\end{equation}
The sum is over all integers $n_1$ and $n_2$, excluding the point $n_1=n_2=0$.
This sum can be calculated numerically, and one finds $I_2=6.6830$, where the contribution of nearest neighbors is $\frac{9}{2}$.
Combining the contributions (\ref{UC3}) and (\ref{UQ5}) one finds
\begin{eqnarray}
\label{app:rcoeff}
 r=\frac{1}{\sqrt{3}}\left(\frac{C}{2{\mathcal K}}-I_2\right)\,,
\end{eqnarray}
where we rescaled the term by $\mathcal K$, as in Eq. \eqref{eq:GL}. Note that $U_{2,Q}$ does not depend on $\theta$, as expected from the general symmetry considerations. This is also the property of the coefficient for the quartic term, as we show now.  The fourth-order contribution to the Coulomb energy is
\begin{equation}
\label{UQQ4}
U_{4,Q}= \frac{{3\mathcal K}}{16} \sum_{i,j} \frac{(z_i-z_{j})^4}{|{\bf r}_i-{\bf r}_j|^5}\,.
\end{equation}
Using similar manipulations as for the second order, one finds
\begin{equation}
\label{UQ54}
U_{4,Q}=\Omega\frac{{9\mathcal K}}{4\sqrt{3}}|\psi|^4 I_4\,,
\end{equation}
with $I_4=3.5596$. After rescaling by $\mathcal K$, this leads to the coefficient
\begin{equation}
\label{ucoeff}
u=\frac{{3\sqrt{3}}}{4} I_4.
\end{equation}
The sixth-order contribution to the Coulomb energy is
\begin{equation}
\label{UQQ46}
U_{6,Q}= -\frac{{5\mathcal K}}{32} \sum_{i,j} \frac{(z_i-z_{j})^6}{|{\bf r}_i-{\bf r}_j|^7}
\end{equation}
Using similar manipulations again one finds
\begin{equation}
\label{UQ56}
U_{6,Q}=\Omega\frac{{5\mathcal K}}{8\sqrt{3}}|\psi|^6 [\cos(6\theta)-10] I_6.
\end{equation}
This is the nonvanishing term at lowest order that depends on $\theta$, as expected from symmetry. From Eq. \eqref{UQ56} we identify the coefficients
\begin{equation}
\label{wcoeff}
w=\frac{5}{8\sqrt{3}} I_6\,,
\end{equation}
and
\begin{equation}
\label{wcoeff}
v=-\frac{25}{4\sqrt{3}} I_6\,,
\end{equation}
where numerical evaluation yields $I_6=2.5577$.

Note that $v$ is negative. This may indicate a first order phase transition.  However, we have found that this is not the case by numerical evaluation of the energy at fixed wave vector ${\bf K}$, as a function of the order parameter $\psi$.    For large confining potential, above the critical confining potential, $C_c=2{\mathcal K} I_2$, the minimum energy lies at $\psi=0$.  At $C_c$, the minimum is still at $\psi=0$, and energy is quartic in $\psi$ at small $\psi$.  Below $C_c$, the minimum energy shifts away from $\psi=0$ and evolves smoothly as a function of $C$.  Hence, there is no first-order phase transition, at least at mean field level, despite the negative value of $v$.  This results from positive higher order terms, such as $|\psi|^8$, which stabilize the second-order phase transition.

In order to calculate the gradient term of Eq.~(\ref{eq:GL}) taking the
contribution of all neighbors into account, we consider  the second-order term of the expansion in the Coulomb interaction for deviation of the wave number $\delta {\bf k}$ from  ${ \bf K}$:
\begin{equation}
\label{fluc} U_{2,Q}(\delta {\bf k})=-{N{\mathcal K} \over
2}\sum_{n_1,n_2}\frac{\sin^2\left(\frac{1}{2}( {\bf K}+\delta {\bf
k})\cdot{\bf r}_{n_1,n_2}\right) }{|{\bf
r}_{n_1,n_2}|^{3}}|\psi|^2\,,
\end{equation}
where ${\bf r}_{n_1,n_2}$ is given by (\ref{coordinate}). Expanding in $\delta {\bf k}$, the
zeroth order term yields Eq.~(\ref{UQ5}), whereas the first order term vanishes in the
sum. The second-order term is
\begin{equation}
\delta U_{2,Q}=-{N{\mathcal K} \over 2}\label{fluc1p}\sum_{n_1,n_2}\frac{\cos({\bf K}\cdot {\bf r}_{n_1,n_2})\left(\delta {\bf k}\cdot{\bf r}_{n_1,n_2}\right)^2 }{4|{\bf r}_{n_1,n_2}|^{3}}|\psi|^2\,,
\end{equation}
which has the form  $\delta U_{2,Q}=M_{ab} \delta k_a\delta k_b$, where the coefficient matrix $M_{ab}$ is symmetric and real.   At the ${\bf K}$ point, the system is invariant under 120 degree rotations.  This implies that the eigenvalues of $M_{ab}$ must be degenerate, {\em i.e.}, $M_{ab}$ must be proportional to the $2\times 2$ identity matrix. Otherwise, the eigenvectors of $M_{ab}$ would pick special directions in the $xy$ plane, thus breaking the rotational symmetry by 120 degrees.  Hence, $M_{xy}=0$ and $M_{xx}=M_{yy}$, allowing us to write $M_{ab}=\frac{M_{xx}+M_{yy}}{2}\delta_{ab}$.  This gives,
\begin{equation}
\label{fluc1}
\delta U_{2,Q}={\Omega{\mathcal K} \over 2} \gamma|\delta {\bf k}|^2|\psi|^2\,,
\end{equation}
with
\begin{eqnarray}
\gamma&=&-\sum_{n_1,n_2} \frac{\cos\left[ {\bf K}\cdot {\bf r}_{n_1,n_2}\right]}{4\sqrt{3}\,|{\bf r}_{n_1,n_2}|}\label{UgradC}\\
&=& 0.2226\,.\label{UgradCnum}
\end{eqnarray}
The resulting gradient term is
\begin{equation}
\label{grad}
U_{\nabla,Q}=\gamma\frac{{\Omega\mathcal K}}{2}|\nabla\psi|^2\,,
\end{equation}
which determines the coefficient $\gamma$ in Eq. \eqref{eq:GL}.

\subsection{Parameters of the GL free energy for the dipolar potential}

\label{UDTALL}

We now calculate the parameters of the Ginzburg Landau free energy for dipolar interactions. Let us assume dipoles of magnitude $P$ are oriented by an external field to point in the $z$ direction. The interaction energy for a pair of dipoles connected by a vector ${\bf r}$ is in units of the lattice spacing
\begin{equation}
\label{UD1}
U'=\frac{{\mathcal K_D}}{|{\bf r}|^3}\left[1-3(\hat{{\bf z}} \cdot \hat{{\bf r}})^2\right]\,,
\end{equation}
where
\begin{equation}
\label{UDD1}
{\mathcal K}_D=\frac{P^2}{a^3}\; .
\end{equation}
while $\hat{{\bf z}}$ and $\hat{{\bf z}}$ are unit vectors in the $z$ and $r$ directions respectively.
For dipoles in the $x-y$ plane then $(\hat{{\bf z}} \cdot \hat{{\bf r}})=0$.
The energy resulting of displacement of dipoles from the $x-y$ plane is
\begin{eqnarray}
\label{UDD}
U_D&=&\frac{{\mathcal K}_D}{2}\sum_{i \ne j}\frac{(x_i-x_j)^2+(y_i-y_j)^2-2(z_i-z_j)^2}{|(x_i-x_j)^2+(y_i-y_j)^2+(z_i-z_j)^2|^{5/2}}\nonumber\\
&&-\frac{\mathcal K}{2} \sum_{i \ne j}\frac{1}{|(x_i-x_j)^2+(y_i-y_j)^2|^{3/2}}\,.
\end{eqnarray}
The vector ${\bf K}$ of long-range order is determined by the nearest neighbors as in the Coulomb case and takes the values of Eqs.~(\ref{eq:K}), (\ref{kmax}). In what follows we expand $U_D$ in powers of $z_i$ and use similar manipulations as in the Coulomb case. The second-order term is
\begin{equation}
\label{UD3}
U_{2,D}= -\frac{ 9 {\mathcal K}_D}{4} \sum_{i,j} \frac{(z_i-z_{j})^2}{|{\bf r}_i-{\bf r}_j|^5}\,.
\end{equation}
Writing the expression in the terms of the order parameter $\psi$ of Eq.~(\ref{eq:heights}) performing a calculation similar to the one resulting in Eq.~(\ref{UQ5}) we find
\begin{equation}
\label{UD5}
U_{2,D}=-\Omega{{9 {\mathcal K}_D} \over \sqrt{3}}|\psi|^2I_2^D\,,
\end{equation}
where
\begin{equation}
\label{InD}
I_n^D=\sum_{n_1,n_2} \frac{\sin^n\left[\frac{\pi}{3}(n_1+n_2)\right]}{|{\bf r}_{n_1,n_2}|^{n+3}}\,,
\end{equation}
with $|{\bf r}_{n_1,n_2}|$ given by (\ref{absr}).
The sum is over all integers $n_1$ and $n_2$,
excluding the point $n_1=n_2=0$.
This sum can be calculated numerically, and one finds $I_2^D=-4.746$.
Combining the contributions (\ref{UC3}) and (\ref{UD5}) one finds
\begin{eqnarray}
\label{app:rcoeffD}
r=\frac{1}{\sqrt{3}}\left(\frac{C}{2{\mathcal K}_D}-9I_2^D\right)\,.
\end{eqnarray}
The fourth-order contribution to the dipolar energy is
\begin{equation}
\label{UDD4}
U_{4,D}= \frac{75{\mathcal K}_D}{16} \sum_{i,j} \frac{(z_i-z_{j})^4}{|{\bf r}_i-{\bf r}_j|^7}\,,
\end{equation}
that gets the form
\begin{equation}
\label{UD54}
U_{4,D}=\Omega\frac{225{\mathcal K}_D}{4\sqrt{3}}|\psi|^4 I_4^D\,,
\end{equation}
with $I_4^D=3.410$.
This leads to
\begin{equation}
\label{ucoeffD}
u=\frac{225}{4\sqrt{3}} I_4^D\,.
\end{equation}
The sixth-order contribution to the dipolar energy is
\begin{equation}
\label{UDD46}
U_{6,D}= -\frac{245{\mathcal K}_D}{32} \sum_{i,j} \frac{(z_i-z_{j})^6}{|{\bf r}_i-{\bf r}_j|^9}\,,
\end{equation}
that gives
\begin{equation}
\label{UD56}
U_{6,D}=\Omega\frac{{245{\mathcal K}_D}}{8\sqrt{3}}|\psi|^6 [\cos(6\theta)-10] I_6^D\,.
\end{equation}
This is the lowest order term that depends on $\theta$, as expected from symmetry, and defines the coefficients
\begin{equation}
\label{wcoeffD}
w=\frac{245}{8\sqrt{3}} I_6^D\,,
\end{equation}
and
\begin{equation}
\label{wcoeffD}
v=-\frac{2450}{8\sqrt{3}} I_6^D\,.
\end{equation}
where numerical evaluation yields $I_6^D=-2.537$.
The gradient term is of a similar form as in Eq.~(\ref{grad}), and its explicit form reads
\begin{eqnarray}
U_{\nabla,D}=\gamma_D \Omega{{9 {\mathcal K}_D} \over 2}|\nabla \psi|^2\,,
\end{eqnarray}
where
\begin{eqnarray}
\gamma_D&=&-\sum_{n_1,n_2} \frac{\cos\left[ {\bf K}\cdot {\bf r}_{n_1,n_2}\right]}{4\sqrt{3}\,|{\bf r}_{n_1,n_2}|^3}\label{UgradD}\\
&=& 0.3366\label{UgradCnumD}\,.
\end{eqnarray}

\section{Effective theory for phase fluctuations}
\label{app:var}

In the $Z_6$-ordered phase, the effective theory describing the
phase fluctuations $\theta({\bf r})$ can be approximated by a massive
quadratic form characterized by a mass $m$. In this Appendix we
derive this effective theory, allowing us to express the physical observables $h$ [Eq.~(\ref{eq:h_def})] and $C({\bf r})$ [Eq.~(\ref{eq:F_k2C(r)})] in terms of $\theta$-correlation functions. To find concrete expressions for these quantities as functions of $T$, we derive
a self-consistent expression for $m$ using a variational
principle. In particular, this provides its
$T$ dependence, generated by fluctuation corrections.

Our starting point is the Euclidean action, which reads
\begin{eqnarray}
\label{exactaction}
S&=&\int_0^{1/T} d\tau\int_\Omega d^2r\,\mathcal{L}\; ,\\
\mathcal{L}&=&\frac{\rho_s}{2}\left\{|\nabla\theta|^2+\frac{1}{c^2}|\partial_\tau\theta|^2\right\}+h_6\cos{6\theta}\nonumber
\end{eqnarray}
where $\tau$ denotes imaginary time and $\Omega$ the spatial area.  Here,  $c$ is the speed of sound, given by Eq.~(\ref{eq:c}), and $\rho_s$ is the renormalized stiffness, which in the ordered phase is approximately equal to the bare stiffness,  Eq.~(\ref{rhos00}). Defining the Fourier
components
\begin{equation}
\theta_q=\int_0^{1/T} d\tau\int d^2r\,e^{-i({\bf k}\cdot{\bf
r}+\omega_n\tau)}\theta({\bf r},\tau)
\end{equation}
where $q\equiv (\omega_n,c {\bf k})$ with $\omega_n=2\pi nT$ the
Matsubara frequencies, the first (free) term in Eq.
(\ref{exactaction}) is recast as
\begin{equation}
\label{Sfree} S_{\rm free}=\frac{T\rho_s}{2c^2}\sum_{\omega_n}\int
\frac{d^2k}{(2\pi)^2}q^2|\theta_q|^2
\end{equation}
while the second (clock) term cannot be written in a simple form.
However, within the ordered phase (dark shaded region in Fig. 2)
this term is relevant and generates a gap, which can be
approximated by a mass term correction to Eq.~(\ref{Sfree}).

We therefore introduce a variational Ansatz for this massive theory
in the form
\begin{equation}
\label{S0_def}
S_0= \frac{T}{2\Omega}\sum_{q,q'}G^{-1}_{q,q'}\delta\theta_q\delta\theta_{q'}
\end{equation}
where the propagator $G_{q,q'}$ assumes the form
\begin{eqnarray}
G_{q,q^\prime}=\frac{T}{\Omega}\langle \delta\theta_q\delta\theta_{q^\prime} \rangle=\delta_{q,-q^\prime}\frac{c^2}{\rho_s(q^2+m^2)}\,,
\label{eq:Gq}
\end{eqnarray}
yielding
\begin{eqnarray}
\label{var_ansatz}
S_0=\frac{T\rho_s}{2c^2}\sum_{\omega_n}\int \frac{d^2k}{(2\pi)^2}(q^2+m^2)|\delta\theta_q|^2\; .
\end{eqnarray}
Here $\delta\theta$ denotes a deviation from the configuration ($\theta_0$) minimizing the clock term, and $m$ is a variational parameter, subsequently adjusted to minimize the free energy. Employing the exact expression
\begin{equation}
F=F_0-T\ln\left[\langle e^{-(S-S_0)}\rangle_0\right]
\end{equation}
where $F_0$ and the expectation value $\langle ...\rangle_0$ are evaluated with respect to $S_0$ [Eq.~(\ref{var_ansatz})], and the convexity relation \cite{Feynman}
\begin{equation}
\langle e^{-(S-S_0)}\rangle \ge e^{-\langle S-S_0\rangle}\; ,
\end{equation}
we find that $F\leq F_{var}$ where
\begin{equation}
\label{F_var}
F_{var}=F_0+T\langle S-S_0\rangle_0\; .
\end{equation}
The minimum of $F_{var}$ with respect to the variational ansatz $S_0$ is therefore an optimal approximation of the free energy $F$.
The condition for a minimum yields
\begin{equation}
\label{F_var_min}
\frac{\partial F_{var}}{\partial G(q)}=\frac{\partial F_{0}}{\partial G(q)}+T\frac{\partial \langle S\rangle_0}{\partial G(q)}=0
\end{equation}
(where $G(q)=G_{q,q}$) for all $q$. Employing Eq.
(\ref{var_ansatz}) we obtain
\begin{eqnarray}
\label{F0fromS0}
F_0&=&-T{\sum_{q}}^\prime\ln[G(q)]\; ,\\
\langle S\rangle_0&=&\frac{T}{\Omega}\left[\frac{1}{2c^2}\sum_q\rho_sq^2G(q)+h_6\langle \cos(6\theta)\rangle_0\right]\; ,\nonumber
\end{eqnarray}
where $\sum_{q}^\prime\equiv \sum _{k_x,k_y}\sum_{\omega_n>0}$, thus avoiding double counting. When substituted in Eq.~(\ref{F_var_min}), one gets a
self-consistent equation for the variational Green's function
$G(q)$:
\begin{equation}
\label{Gq_min}
G^{-1}(q)=\frac{\rho_s q^2}{c^2}+2h_6\frac{\partial \langle \cos(6\theta)\rangle_0}{\partial G(q)}\; .
\end{equation}
Recalling that $\theta=\theta_0+\delta\theta$ where $\theta_0$ is
one of the minima of the clock term [{\em i.e.} $\langle
\cos(6\theta_0)\rangle_0=-{\rm Sign}(h_6)$], and using the
quadratic form of $S_0$, the expectation value $\langle
\cos(\beta\theta)\rangle_0$ for arbitrary $\beta$ can be expressed as
\begin{eqnarray}
\langle \cos(\beta\delta\theta) \rangle &= &e^{-\frac{\beta^2}{2}\langle (\delta\theta)^2 \rangle}\; ,\nonumber \\
\langle(\delta\theta)^2\rangle &=&\frac{T}{\Omega}\sum_q G(q)\; .
\label{eq:gaussian_theta}
\end{eqnarray}
For $\beta=6$, this yields
\begin{equation}
\label{cosine_Gq}
\langle \cos(6\theta)\rangle_0=-{\rm Sign}(h_6)e^{-18\langle(\delta\theta)^2\rangle}\; .
\end{equation}
Finally, employing $G^{-1}(q)=\rho_s(q^2+m^2)/c^2$ [Eq.~(\ref{eq:Gq})] and inserting Eq.~(\ref{cosine_Gq}) in Eq.~(\ref{Gq_min}),
we arrive at a self-consistent equation for the mass $m$:
\begin{eqnarray}
\label{m_selfconsistent}
m^2 &=& \frac{36c^2 |h_6|}{\rho_s}\exp\left\{-18\langle(\delta\theta)^2\rangle \}\right\}\; ,  \\
\langle(\delta\theta)^2\rangle &=&\frac{T}{\Omega}\sum_q\frac{c^2}{\rho_s(q^2+m^2)} \; . \nonumber
\end{eqnarray}
The last sum can be evaluated for arbitrary $T$:
\begin{eqnarray}
\label{eq:theta_corr_0}
\langle(\delta\theta)^2\rangle &=& \frac{c^2}{4\pi\rho_s}\int_0^\Lambda\frac{dk\,k}{\sqrt{c^2k^2+m^2}}\coth\left(\frac{\sqrt{c^2k^2+m^2}}{2T}\right)\nonumber\,,\\
&=& \eta(T)\left\{\ln\sinh\left(\frac{c\Lambda}{2T}\right)-\ln\sinh\left(\frac{m}{2T}\right)\right\}
\end{eqnarray}
where $\eta(T)$ is given by Eq.~\eqref{eq:etarhos}, and $\Lambda$ is the ultraviolet (UV) cutoff. Inserting this expression (using $c\Lambda\gg T$) into the first line of Eq.~(\ref{eq:gaussian_theta}) with $\beta=1$, and employing Eq.~(\ref{eq:h_def}), we thus
obtain the expression Eq.~(\ref{eq:hvsT}) for the layer separation $h$ as a
function of $T$ for arbitrary $T<T_6$ and $r<r_c$. Similarly, substitution of Eq.~(\ref{eq:theta_corr_0}) in Eq.~(\ref{m_selfconsistent}) results in
\begin{equation}
\label{m_vs_T}
m^2=m_0^2\exp\left\{18\eta(T)\ln[2\sinh(m/2T)]\right\}
\end{equation}
where
\begin{equation}
\label{m0def}
m_0^2=(36|h_6| c^2/\rho_s)e^{-9c\Lambda/2\pi\rho_s}
\end{equation}
is $T$-independent. The self-consistent solution of Eq.~(\ref{m_vs_T}) yields the $T$-dependent mass $m(T)$.
In the low $T$ regime $T\ll m$ (below the
dashed line in Fig. 2), one obtains a constant and $m\approx m_0$.
However in the classically ordered regime $m\ll T \ll T_6$ we get
\begin{equation}
\label{m_vs_T_class}
m(T)\approx T\left(\frac{m_0}{T}\right)^{\frac{T_6}{T_6-T}}
\end{equation}
where we have used $T_6=(2\pi\rho_s/9)$ (see Sec.~\ref{sec:pd}A).

\section{Evaluation of the structure form factor}
\label{App:C}

To obtain the diffraction intensity $I_{{\bf k}_{3d}}$
from Eq. \eqref{eq:I2S_k}, we recall the definition of the structure factor
\begin{eqnarray}
S_{{\bf k}_{3d}}=\sum_{i}e^{i({\bf k}\cdot{\bf r_i}+k_zz_i)}\; .
\label{eq:S_k_def}
\end{eqnarray}
Inserting into Eq. \eqref{eq:I2S_k}, one obtains
\begin{eqnarray}
I_{{\bf k}_{3d}}= \sum_{i,j}e^{i{\bf k}\cdot({\bf r_i}-{\bf
r_j})}\langle e^{ik_z(z_i-z_j)} \rangle\; . \label{eq:I_k}
\end{eqnarray}
For $k_zh\ll 1$, the last exponent can be expanded, yielding
\begin{eqnarray}
I_{{\bf k}_{3d}}=I_{({\bf k},k_z=0)}+F({\bf k},k_z)\,,
\end{eqnarray}
with the leading $k_z$-dependent contribution
\begin{eqnarray}
F({\bf k},k_z)= k_z^2\sum_{i,j}e^{i{\bf k}\cdot({\bf r_i}-{\bf r_j})}\langle z_iz_j \rangle \label{eq:F_k}
\end{eqnarray}
where we have omitted a term $\propto\delta({\bf k})$. Employing
Eq.~(\ref{eq:heights}), we express the heights $z_i$ as
\begin{eqnarray}
z_i=\frac{|\psi_0|}{2}\left(e^{i(\theta_0+{\bf K}\cdot{\bf r_i})}e^{i\delta\theta({\bf r_i})}+e^{-i(\theta_0+{\bf K}\cdot{\bf r_i})}e^{-i\delta\theta({\bf r_i})} \right)
\label{eq:z_i2theta}
\end{eqnarray}
where ${\bf K}$ is given by Eq.~(\ref{eq:K}).
When substituted in Eq.~(\ref{eq:F_k}), one obtains a summation
over terms with phase-factors $e^{\pm i{\bf K}\cdot({\bf r_i}+{\bf
r_j})}$ or $e^{\pm i{\bf K}\cdot({\bf r_i}-{\bf r_j})}$.
Performing the sum over the center-of-mass coordinates first, the
former type yields zero; from the latter type we get $F({\bf k},k_z)$ of the form Eq.~(\ref{eq:F_k2C(r)}), which relates it to the correlation function $C({\bf r})$.

We now evaluate $C({\bf r})$ within the Gaussian theory for the fluctuating field $\delta\theta$ derived in Appendix \ref{app:var}, which relates it to the correlation function $\langle \delta\theta({\bf r})\delta\theta(0) \rangle$:
\begin{eqnarray}
C({\bf r})=  \langle \cos(\delta\theta) \rangle^2e^{\langle \delta\theta({\bf r})\delta\theta(0) \rangle}
\label{eq:gaussian}
\end{eqnarray}
where we have used Eq.~(\ref{eq:gaussian_theta}) for $\langle \cos(\delta\theta) \rangle$.
Inside the ordered phase, the Gaussian theory for $\delta\theta$
is massive and the propagator in momentum space is given by Eq.~(\ref{eq:Gq})
with $m$ the mass associated with the clock-ordering term, evaluated as a function of $T$
and the parameters of the model using a variational principle in
Appendix \ref{app:var}. We thus obtain
\begin{eqnarray}
\label{eq:theta_corr}
\langle \delta\theta({\bf r})\delta\theta(0) \rangle&=&\frac{T}{\Omega}\sum_{q}e^{i{\bf k}\cdot{\bf r}}G_{q,q} \\
&=
&\frac{c^2}{4\pi\rho_s}\int_0^\Lambda\frac{dk\,kJ_0\left(k|{\bf r}|\right)}{\sqrt{c^2k^2+m^2}}\coth\left(\frac{\sqrt{c^2k^2+m^2}}{2T}\right)\nonumber\,.
\end{eqnarray}
From this general expression, one can obtain asymptotic expressions for $\langle
\delta\theta({\bf r})\delta\theta(0) \rangle$ and consequently
$C({\bf r})$ for large $|{\bf r}|$ in two limits - for $T$ above or below
the crossover line:
\begin{eqnarray}
\label{eq:theta_corr_T}
\langle \delta\theta({\bf r})\delta\theta(0) \rangle &\approx & \frac{c}{4\pi\rho_s |{\bf r}|}e^{-m|{\bf r}|/c}\quad (T\ll m) \\
\langle \delta\theta({\bf r})\delta\theta(0) \rangle &\approx &
\eta(T)K_0\left(\frac{m|{\bf r}|}{c}\right)\quad (m\ll T<T_{6})\,,\nonumber
\end{eqnarray}
with $K_0(z)$ the modified Bessel function. Inserting in Eq.
(\ref{eq:gaussian}), this yields corresponding expressions for $C({\bf r})$. The scale of exponential decay in both expressions, $c/m$, dictates the width of the Bragg peaks.

\end{document}